# Impact Craters on Pluto and Charon and Terrain Age Estimates

**K. N. Singer**
*Southwest Research Institute*

**S. Greenstreet**
*University of Washington*

**P. M. Schenk**
*Lunar and Planetary Institute*

**S. J. Robbins**
*Southwest Research Institute*

**V. J. Bray**
*University of Arizona*

Pluto's terrains display a diversity of crater retention ages ranging from areas with no identifiable craters to heavily cratered terrains. This variation in crater densities is consistent with geologic activity occurring throughout Pluto's history and also a variety of resurfacing styles, including both exogenic and endogenic processes. Using estimates of impact flux and cratering rates over time, Pluto's heavily cratered terrains appear to be relatively ancient, 4 Ga or older. Charon's smooth plains, informally named Vulcan Planitia, did experience early resurfacing, but there is a relatively high spatial density of craters on Vulcan Planitia and almost all overprint the other types of volcanic or tectonic features. Both Vulcan Planitia and the northern terrains on Charon are also estimated to be ancient, 4 Ga or older. The craters on Pluto and Charon also show a distinct break in their size-frequency distributions (SFDs), where craters smaller than ~10-15 km in diameter have a shallower SFD power-law slope than those larger than this break diameter. This SFD shape on Pluto and Charon is different than what is observed on the Earth's Moon, and gives the Kuiper belt impactor SFD a different shape than that of the asteroid belt.

## 1. INTRODUCTION

One of the goals of the New Horizons mission was to use impact craters observed on Pluto and Charon to both understand geologic surface processes and learn about the size-distribution of the greater population of Kuiper belt objects (*Stern,* 2008; *Young et al.,* 2008). It was not known if there would be many impact craters on Pluto, or if craters would be resurfaced and effectively erased by atmospheric and surface processes.  Charon, however, was expected to have craters.  Craters were discovered on both worlds and provided a wealth of knowledge about the surface ages and geologic processes operating on Pluto and Charon.  Additionally, the impact craters on Pluto and Charon yielded insights into the population of impactors in the outer solar system and the origin and evolution of planetesimals.

Craters are useful geologic tools for investigating a number of topics.  Because the initial shape of a crater of a given size is fairly well known for many surface materials, deviations from that initial form can reveal information either about the crater formation conditions (e.g., formation in a thin shell) or the geologic processes that later modified a crater.  Both the depth and morphology of modified impacts can be compared to the same properties for relatively fresh or unmodified impacts to produce a qualitative and/or quantitative estimate of the processes occurring over time.  Craters also excavate the surface, allowing a view of the near-surface interior.  Both the walls and ejecta deposits of craters can reveal subsurface layering and materials that may not be detectable on the surface.  Ejected material, being a thinner deposit, erodes more quickly, and thus can serve as a time marker sensitive to relatively recent epochs.  Larger craters form more frequently in early solar system history, and their large size often allows them to persist longer under erosive processes, thus they can be a witness to the more distant past and longer timescales.

Because craters are thought to have formed with a relatively steady rate over the last ~4 billion years (there are exceptions to this), the variation in the spatial density of craters between different geologic units speaks to their relative age differences.  Younger terrains have fewer craters, and older terrains generally have higher crater densities and more large craters.  Crater densities combined with models of the impact flux and cratering rates over time can give an estimate of the age of the surface in units of time, usually expressed in millions (Myr) or billions (Gyr) of years.

Through scaling laws, the craters can also be related to the impactors that made them.  The size distribution of the impacting population can be derived.  This requires taking into account any resurfacing that may have occurred once the craters formed, and accounting for any secondary crater or circumplanetary populations from debris in the system, either created by fragments ejected from primary impacts or from breaking up of small moons.  Having multiple terrains—and in the case of the Pluto-system, multiple bodies—to compare among can help discriminate between a primary impacting population and other populations/geological effects.

Here we describe the above type of investigations performed using New Horizons data of Pluto and Charon.  Note some feature names used in this chapter are formal and others are informal.  Please refer to the Nomenclature Appendix in this volume).  The image sequences or scan names used here refer to the Request ID for each observation, which is a unique identifier that can be found in the headers of the data as archived on the official repository for New Horizons data, the Planetary Data System (PDS) Small Bodies Node (https://pds-smallbodies.astro.umd.edu/data_sb/missions/newhorizons/index.shtml).  The images come from two instruments on New Horizons: the LOng Range Reconnaissance Imager (LORRI) and the Multispectral Visible Imaging Camera (MVIC). Details about the relevant datasets can be found in **Table 1.1**. New Horizons flew through the Pluto-system and thus observed one side of Pluto better than the other. We define the encounter hemisphere of both Pluto and Charon as the surface area of each body observed at higher resolution (pixels scales of ~76–865 m px$^{-1}$) during the closest approach of the spacecraft during its flyby (**Fig. 3.1**). The non-encounter hemisphere consists of portions of Pluto and Charon that were seen at lower resolution (pixels scales of ~2.2–32 km px$^{-1}$) during New Horizons' approach to the system. One Pluto or Charon day is 6.4 Earth days. As they rotated, New Horizons was able to see different portions of the bodies on approach, hence the wide range of resolutions for the non-encounter hemisphere.

**Table 1.1.** Modified from *Singer et al.,* 2019.

| Request ID | Plot Legend Short Title | Instrument Mode | ~Pixel Scale [m px$^{-1}$] | Mosaic size or Scan | Exposure or Scan rate |
|---|---|---|---|---|---|
| Pluto | | | | | |
| PELR_P_LORRI | | 1×1 | 850 ± 30 | 4×5 | 150 ms |
| PELR_P_LEISA_HIRES‡ | | 1×1 | 234 ± 13 | 1×12 | 50 ms |
| PELR_P_MPAN_1‡ | | 1×1 | 117 ± 2 | 1×27 | 10 ms |
| PELR_P_ MVIC_ LORRI_CA‡ | LCA | 1×1 | 76 | 1×35 | 10 ms |
| PEMV_P_MPAN1 | MPAN | Pan TDI 1 | 480 ± 5 | Scan | 1600 μrad s$^{-1}$ |
| PEMV_P_ MVIC_ LORRI_CA | MCA | Pan TDI 2 | 315 ± 8 | Scan | 1000 μrad s$^{-1}$ |
| Charon | | | | | |
| PELR_C_LORRI | C_LORRI | 1×1 | 865 ± 10 | 2×4 | 150 ms |
| PELR_C_LEISA_HIRES‡ | C_LEHI | 1×1 | 410 ± 5 | 1×7 | 60 ms |
| PELR_C_ MVIC_ LORRI_CA‡ | LORRI_CA | 1×1 | 154 | 1×9 | 10 ms |
| PEMV_C_ MVIC_ LORRI_CA | MVIC_CA | Pan TDI 1 | 622 | Scan | 1000 μrad s$^{-1}$ |

## 2. CRATER MORPHOLOGIES

The final morphology of a pristine impact crater is a result of the collapse of a geometrically simple, bowl-shaped 'transient crater' that forms shortly after impact from a combination of excavation and compression of the surface (e.g., *Melosh,* 1989). For simple craters this involves minor rim debris sliding, but for complex craters, floor uplift and rim failure are involved. This collapse is controlled by a complex interplay of crustal material strength, target body gravity, and impactor energy. Although ice in many outer solar system planetary settings (at temperatures of tens to 100 Kelvin) is significantly stronger than ice on the surface of the Earth, which is much closer to its melting point of 273 K (e.g., *Durham and Stern,* 2001), it is still weaker than rock in terms of tensile and compressive strength (also see chapter by *Umurhan et al.* in this volume). On account of this basic strength difference, the amount of collapse of the transient cavity during crater modification is greater and occurs at smaller diameters ($D$) for craters forming in ice than on a rocky body of similar gravity (e.g., *Schenk et al.,* 2004).

Craters on Pluto and Charon display many similarities to craters on other icy worlds, including a transition from smaller, bowl-shaped craters referred to as simple craters, to larger, flatter (e.g., more pie-pan shaped) craters, some with central peaks (**Fig. 2.1**; also see figures in Section 3). There is one crater on Pluto with a deep central depression (an eroded 85-km crater at 5.7°S, 155.3°E; **Fig. 2.1e-h**), that appears similar to central pit craters seen on icy satellites such as Ganymede and also on Mars and Ceres (e.g., *Schenk,* 1993; *Alzate and Barlow,* 2011; *Bray et al.,* 2012; *Conrad et al.,* 2019). The expected transition diameter from central peak to central pit craters on Pluto, assuming $g^{-1}$ scaling from Ganymede and Callisto, is ~55 km. Thus, the lack of prevalence for central pit morphologies at larger diameters is somewhat surprising. However, there are only a handful of craters larger than 55 km in diameter and smaller than the very largest basins, so this limits the possible examples.

The largest impact feature seen on Pluto is the very large ($D$~800-1000 km) Sputnik basin that is partly filled by the nitrogen-rich ice sheet of Sputnik Planitia; the second largest is Burney basin (~240 km in diameter). Burney basin lies just north of Sputnik Planitia and exhibits multiple ring-structures, the only structure confirmed to do so (*Moore et al.,* 2016). Both Sputnik and Burney are extensively eroded and modified. Burney basin preserves a depth of 2-3 km (*Schenk et al.,* 2018c). The surface of Sputnik Planitia is 2.5-3.5 km below the eroded edge of the basin, but the initial, unfilled depth of Sputnik basin may have been as deep as 10 km (*McKinnon et al.,* 2016). The circular Simonelli feature (**Fig. 2.1d**) on the non-encounter hemisphere of Pluto (which was only seen at lower resolution) is similar in size to Burney and appears to show a concentric, ring-like structure, and potentially also a large central peak. Some of Simonelli's appearance is due to topography, and some may be due to deposits of bright ice in topographic lows emphasizing the concentric nature (*Stern et al.,* 2020).

After a crater forms, geologic processes can reshape or erode the crater over time and most craters on Pluto show signs of at least some modification. Geologic processes acting on Pluto are described in detail in many sources (e.g., *Moore et al.* 2016; also see chapter by *White et al.* in this volume, and chapter by *Moore and Howard* in this volume) and those processes specific to crater modification are discussed in the main text and supplement of *Singer et al.*, (2019). These processes act in some areas to erode craters while in other areas craters can be either mantled or infilled (e.g., a few craters have nitrogen-rich ice deposits on their floors similar to the plains of Sputnik Planitia). Erosion of ejecta deposits appears to occur quickly on Pluto as very few can be easily identified. The processes affecting individual regions are discussed in Section 3.

Craters on Charon also generally exhibit the progression from bowl-shaped to complex morphologies with increasing size expected for an icy body (**Fig 2.2.;** also see figures in Section 3). The craters on Charon can serve as a good reference for what a more pristine crater would look like on Pluto, although the gravity is lower on Charon ($g$ ~0.3 m s$^{-2}$) than on Pluto ($g$ ~0.6 m s$^{-2}$) and this difference must be taken into account when comparing crater morphologies between the two bodies.  On Charon many of the craters larger than $D$~10 km have extensively ridged crater floors, similar to those seen on icy saturnian satellites (*White et al.,* 2013; *White et al.,* 2017; *Schenk et al.,* 2018b; *Schenk et al.,* 2018a) and on ice-rich Ceres (*Schenk et al.,* 2019) , and are diagnostic of floor uplift and deformation.  Depths of relatively unmodified complex craters on Pluto and Charon have been measured (*Schenk et al.,* 2018c; *Schenk et al.,* 2018b).  These depths indicate simple-to-complex transition diameters (from an inflection in a plot crater depth vs diameter) of ~4.3 km and ~5.3 km on Pluto and Charon, respectively.  The morphological transition to central peaks occurs at $D$ ~12.5 km on Charon.  These transition diameters and the range of depths on the two bodies are consistent with gravity scaling of crater dimensions in hypervelocity craters on ice-rich targets (*Schenk et al.,* 2018b).

    Ejecta deposits around Charon craters take a variety of forms.  Many craters with a distinct albedo pattern that is a combination of dark inner ejecta and bright outer ejecta/rays are found in Oz Terra where the overhead lighting is well suited for observing albedo variation (*Robbins et al.,* 2019).  Some craters on Vulcan Planitia (**Fig. 2.2**) also display thicker ejecta deposits with distinct margins similar to those found on other icy bodies (*Robbins et al.,* 2018), including Ganymede and Dione (*Horner and Greeley,* 1982; *Schenk et al.,* 2004; *Boyce et al.,* 2010; *Schenk et al.,* 2018a) as well as Mars (*Mouginis-Mark,* 1979; *Costard,* 1989; *Barlow and Bradley,* 1990; *Barlow et al.,* 2000; *Robbins et al.,* 2018).  These thicker ejecta deposits are identified in areas (such as on Vulcan Planitia) where the lighting is oblique and enables the topography of more subtle features to be seen.  Thus, the fact that they are observed more on Vulcan Planitia may be due to lighting geometry effects.  Secondary cratering (or the lack thereof) is discussed in Section 5.1 below.

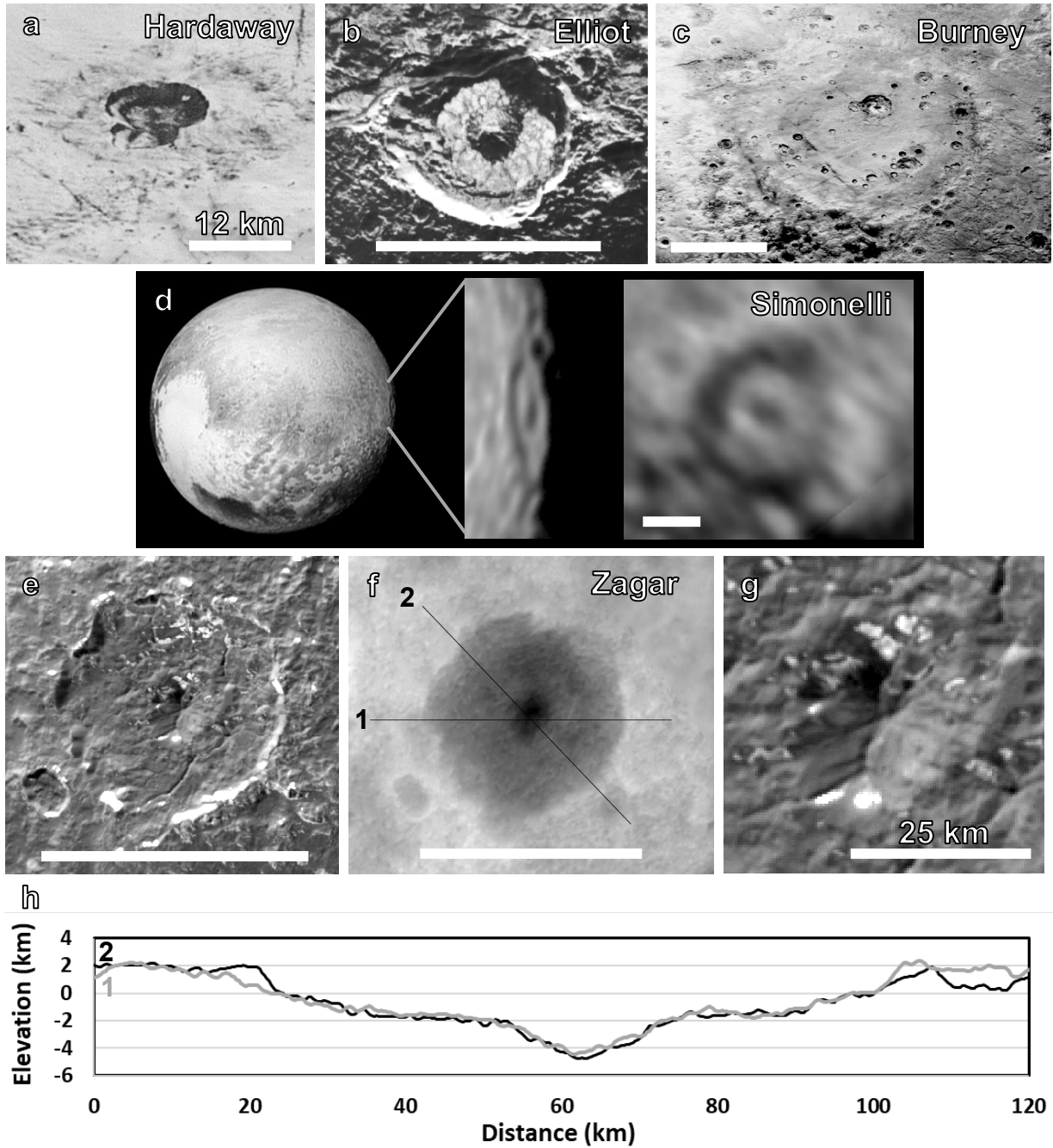

**Figure 2.1** Examples of Pluto crater morphologies including: (a) Portion of the highest resolution image strip showing a double crater with distinct layering in the walls and hints of dark material in an ejecta deposit (this is one of only a few examples where ejecta deposits can be easily detected on Pluto), (b) central peak with wall terraces and nitrogen-rich ice deposits in the floor, (c) basin with multiple ring structures, (d) three views of a possible multi-ring structure with a large central peak on the non-encounter hemisphere of Pluto, (e) crater with deep central pit, (f) topography of central pit crater and surroundings, where white is high and black is low, and the elevations range from approximately +3.8 km to -5.3 km (a linear stretch is applied), (g) close-up of central pit structure, and (h) topographic profiles as shown in panel f (the profiles

start at the numbered side). All scale bars are 100 km except where noted. Image sources: a from PELR_P_MVIC_LORRI_CA (76 m px$^{-1}$); b, c, e, g from PEMV_P_MVIC_LORRI_CA (315 m px$^{-1}$); d from PELR_PC_MULTI_MAP_B_12_L1AH (16.7 km px$^{-1}$), where the right-most panel is a simple cylindrical re-projection. Note that in this figure and subsequent figures, different stretches of the pixel values have been applied to different panels, therefore absolute brightness cannot be compared across the frames.

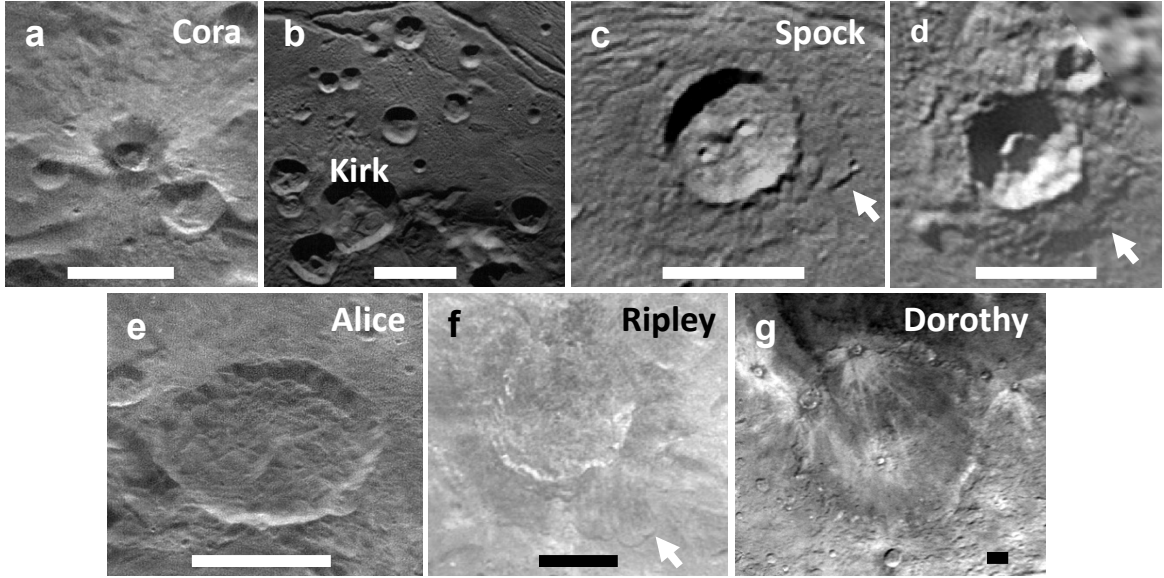

**Figure 2.2** Examples of Charon crater morphologies including: (a) distinctive ejecta albedo patterns, (b) smaller bowl-shaped simple craters and craters with mass wasted material on their floors, (c-d) complex craters with central peaks and thick ejecta deposits, (e-f) complex craters with heavily terraced floors, and (g) the largest identifiable crater on Charon's encounter hemisphere (~250 km in diameter). All scale bars are 30 km in length. White arrows indicate the margins of thicker ejecta deposits and in the case of panel f, a landslide-like ejecta deposit. Image sources: a-b,e from PELR_C_MVIC_LORRI_CA (160 m px$^{-1}$), c-d from PEMV_C_MVIC_LORRI_CA (622 m px$^{-1}$); and f-g from PELR_C_LORRI (865 m px$^{-1}$).

## 3. REGIONAL CRATER SIZE-FREQUENCY DISTRIBUTIONS

Impact crater size-frequency distributions (SFDs) are useful for understanding both the geologic histories of Pluto and Charon and the impactor populations that formed the craters (*Moore et al.,* 2016; *Robbins et al.,* 2017; *Singer et al.,* 2019). The plots below are displayed in an R-plot format (*Crater Analysis Techniques Working Group*, 1979), where the "R" stands for relative. The differential number of craters for a given diameter (*D*) bin is proportional to a power law with an exponent of $q$ ($dN/dD \propto D^q$), where $q$ is often referred to as the log-log slope. The R-plot SFD divides this differential SFD by $D^{-3}$ such that a differential distribution with $q = -3$ is a horizontal line, and $q = -4$ and $-2$ form lines that slope downward and upward with increasing $D$, respectively (see guide in **Fig. 3.6b**). Because a slope of $q = -3$ is commonly seen, the R-plot helps visually distinguish changes from this slope as a function of diameter and between different crater populations. For each distribution, we normalize the number of craters per bin by the mapped area to give the density of craters per size bin. In this section, we also follow the convention of making all of the plots square (e.g., one order of magnitude on the x-axis is the same physical length as 1 order of magnitude on the y-axis) so that the slopes can be directly visually compared between different plots. The data and areas for the distributions is described in *Singer et al.* (2019) and the full dataset is also archived under the following DOI: 10.6084/m9.figshare.11904786.

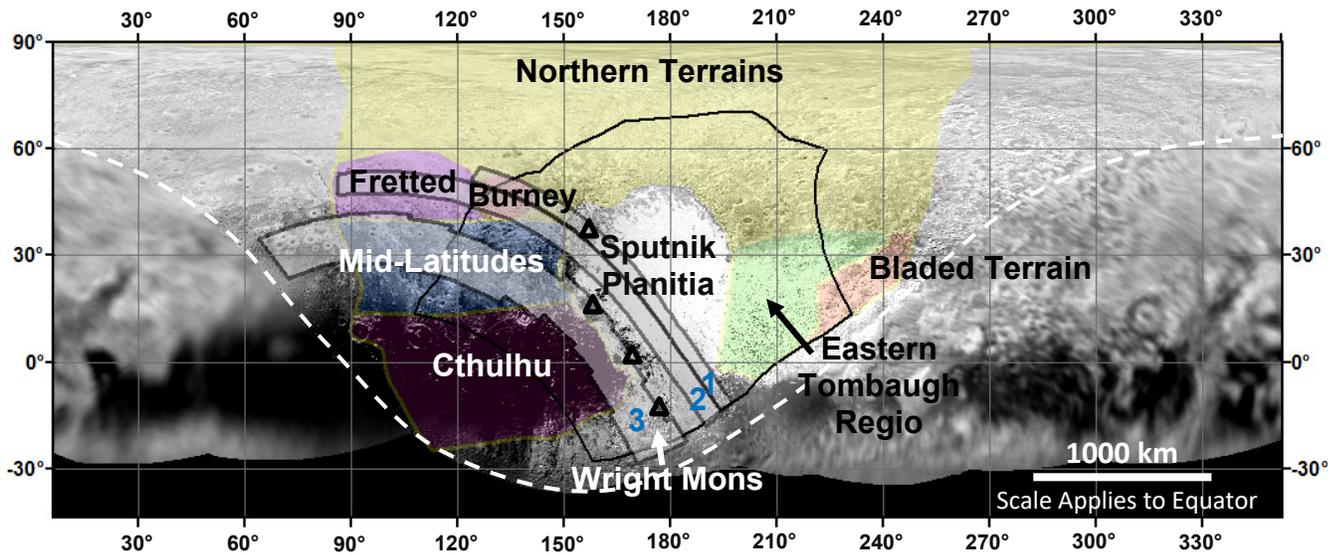

**Figure 3.1** Simple cylindrical projection of the Pluto basemap mosaic available on the PDS (https://pds-smallbodies.astro.umd.edu/holdings/nh-p_psa-lorri_mvic-5-geophys-v1.0/dataset.shtml) with the shaded areas demarcating the broad regions described in the text. See Sections 3 and 5 for details of each region. Colored areas indicate extent of mappable PEMV_P_MPAN1 (480 m px$^{-1}$) and solid black outline encompasses the more limited area from PEMV_P_MVIC_LORRI_CA (315 m px$^{-1}$). Small black triangles indicate the chaotic mountain block regions. The high resolution strips are also shown and labeled 1, 2, and 3 in blue text for PELR_P_MVIC_LORRI_CA (76 m px$^{-1}$), PELR_P_MPAN_1 (117 m px$^{-1}$), and PELR_P_LEISA_HiRES (234 m px$^{-1}$), respectively. White dashed curve separates the encounter hemisphere from the non-encounter hemisphere. The black regions at the bottom of the map were regions in an extended period of darkness (Pluto night) during the time of the New Horizons flyby. Modified from *Singer et al.* (2019).

## 3.1. Pluto Terrains

Here we describe the broad physiographic provinces on Pluto and present their crater size-frequency distributions for terrain age analysis. We briefly comment on the general geologic context of the regions, especially as related to degradation of craters, or superposition relationships of other geologic processes compared with cratering. More details on the geologic context for different regions on Pluto can be found in the chapter by *White et al.* in this volume.

*3.1.1. Cthulhu Macula* Pluto's dark equatorial band has as complicated and diverse a geologic history as any other location on Pluto. Although Cthulhu Macula is unified in its dark albedo, likely a tholin coating of variable thickness (*Cruikshank et al.,* 2015; *Stern et al.,* 2015; *Grundy et al.,* 2016; *Protopapa et al.,* 2017; *Schmitt et al.,* 2017; *Grundy et al.,* 2018; *Cook et al.,* 2019), terrains vary from heavily cratered, apparently ancient regions, to smooth, more lightly cratered plains (**Fig. 3.2**). Here we use the entire broad physiographic province of Cthulhu (**Fig. 3.1**) on the crater plots (**Fig. 3.2d**), meaning that this represents an average crater spatial density for the entire region. The two largest craters after Burney basin ($D \approx 240$ km), Edgeworth ($D \approx 140$ km) and Oort ($D \approx 110$ km), are located in Cthulhu. The existence of many large craters in Cthulhu with varying preservation states indicates a relatively ancient surface overall. In addition to higher resolution, MVIC instrument scans of Cthulhu have better signal to noise than the LORRI instrument coverage, allowing a more detailed look at this dark terrain.

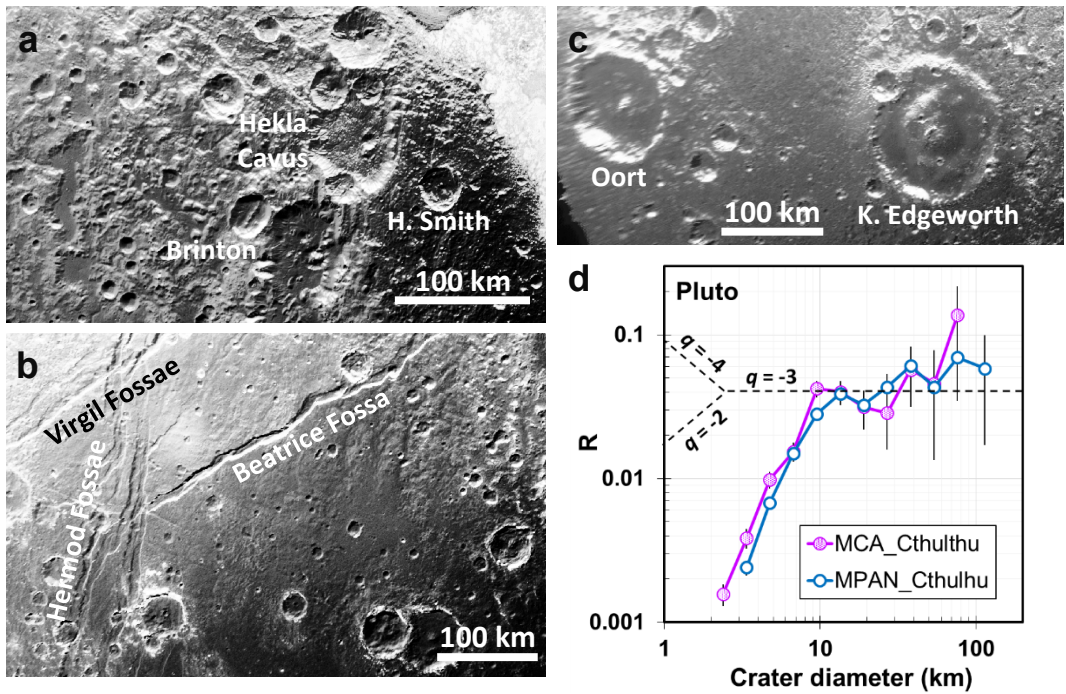

**Figure 3.2** Surfaces of Cthulhu range from (a) heavily cratered and rugged to (b) smoother and lightly cratered. (c) Two large craters on the encounter hemisphere. (d) R-plot of the Cthulhu region in two different datasets (see **Fig. 3.1** for dataset extents and *Singer et al.* 2019 for details). Also shown in panel d are dashed lines representing a differential slope of -4, -3, and -2 for reference. Image sources: a from PEMV_P_MVIC_LORRI_CA (315 m px$^{-1}$) noted as MCA in the legend of this plot and subsequent plots; b and c from PEMV_P_MPAN1 (480 m px$^{-1}$) noted as MPAN in plot legends.

***3.1.2. Western mid-latitudes*** The mid-latitude region described here lies north of Cthulhu and represents a zone of transitional albedo between the dark Cthulhu and the brighter northern terrains (**Fig. 3.3**). This region contains both relatively heavily cratered areas and large eroded plains (i.e. Piri and Bird Planitiae) that show few craters (*Moore et al.,* 2016; *Moore et al.,* 2017). The latitude of 40° was selected as the northern boundary, but the albedo transition is gradual. Distinctive "bright-halo" craters with methane deposits (*Grundy et al.,* 2016) exist both here and in the fretted terrain to the north (see next section).

The mid-latitudes exhibit the same paucity of small craters ($D < 10$ km) as seen on other terrains on Pluto (and Charon as described at length in section 3.3), but are also deficient in the largest craters compared to Cthulhu. The surface area of Cthulhu seen in the MVIC close approach scan (MCA_Cthulhu in **Figs. 3.2d**) is similar to that of the mid-latitude region seen in the same scan (MCA_Mid-lat in **Figs. 3.3b**) and the Cthulhu region has 4 craters larger than $D$=50 km, while the mid-latitude region has none.

It could be by chance that not many large craters occur in this region, but also significant resurfacing, possibly via an early mantling episode, may have erased some large craters in this region. The lack of large craters suggest early (rather than later) resurfacing, because large craters should occur more frequently in Pluto's early history. Eroded scarps (notably Piri Rupes) and bright-halo craters suggest sublimation and redeposition of frosts have played a large role in modifying this region over various timescales, possibly related to Pluto's different seasonal zones (*Binzel et al.,* 2017; *Earle et al.,* 2017). The exact configuration of large smooth regions adjacent to heavily cratered regions is not easily explained, however. Overall this region appears middle-aged to old, but large areas have been, and possibly continue to be, resurfaced.

***3.1.3. Fretted terrain, Burney Basin, and other northern terrains*** North of the 40$^{th}$ parallel Pluto has yet another diverse set of terrains (**Fig. 3.4**). Starting in the west, a set of unique conspicuous eroded valleys termed "fretted terrain" (*Howard et al.,* 2017; also see chapter by Moore and Howard in this volume) cover an approximately 250x450 km area. Although there are very few, if any, cases where the valleys can be seen to directly cut through or breach crater rims, in some areas the valleys appear to be diverted around crater rims. This pattern indicates the crust itself is old and the valleys are a later feature, consistent with some of the craters' degraded appearance.

Burney basin lies to the east of the fretted terrain and represents a unique surface. The highest resolution LORRI strip (pixel scale of 76 m px$^{-1}$) passes over Burney. The area inside the basin rim is generally smoother than much of the surrounding terrain. There are many possible reasons for this. The interior of the basin may erode differently than its surroundings due to the basin formation altering the material properties in this location. Alternately, or in addition, the basin may collect more atmospheric deposits because of its bowl-like geometry. A more thorough future investigation involving modeling and topographic data may shed light on this topic. Burney basin is also a site of so-called washboard terrain which has been suggested to be an ancient texture, fitting with the likely old age of the basin (*White et al.,* 2019).

    Nitrogen ice ($N_2$), methane ice ($CH_4$), and carbon monoxide (CO) sublimate and redeposit in Pluto's seasonal and mega-seasonal cycles (e.g., *Stansberry,* 1994; *Spencer et al.,* 1997; *Young,* 2013; *Bertrand et al.,* 2018; *Bertrand et al.,* 2019; also see chapter by Young et al. in this volume), and radiolytic processing of $CH_4$ either in the atmosphere or on the surface creates heavier, darker, and generally redder complex molecules over time, which have been referred to as haze particles or tholins (e.g., *Grundy et al.,* 2018). To this point, several alternating bright and dark layers are seen in a few craters in the highest resolution strip (**Figs. 2.1a and 3.4c**) hinting at previous epochs of deposition. These alternating dark/bright layers do not look like the product of mass wasting alone: the layers are seen at approximately the same elevation in several craters, and also in fault and mountain block walls seen elsewhere on Pluto (*Moore et al.,* 2016). This image strip also reveals dark ejecta around the freshest crater (Hardaway crater in **Figs. 2.1a and 3.4c**), and possible dark ejecta blocks around a few others. The older, more degraded craters show hints of dark material, but it appears that bright volatile deposition has occurred over the entire region, covering any darker material ejected by the older craters. Therefore, the general albedo of craters in this region gives an indication of relative age for a given impact. With a refined cratering rate model, this information might put constraints on the rate of volatile deposition in this region. It should be noted, however, that different regions on Pluto do not follow this same pattern (e.g., in Cthulhu where the surface is primarily dark).

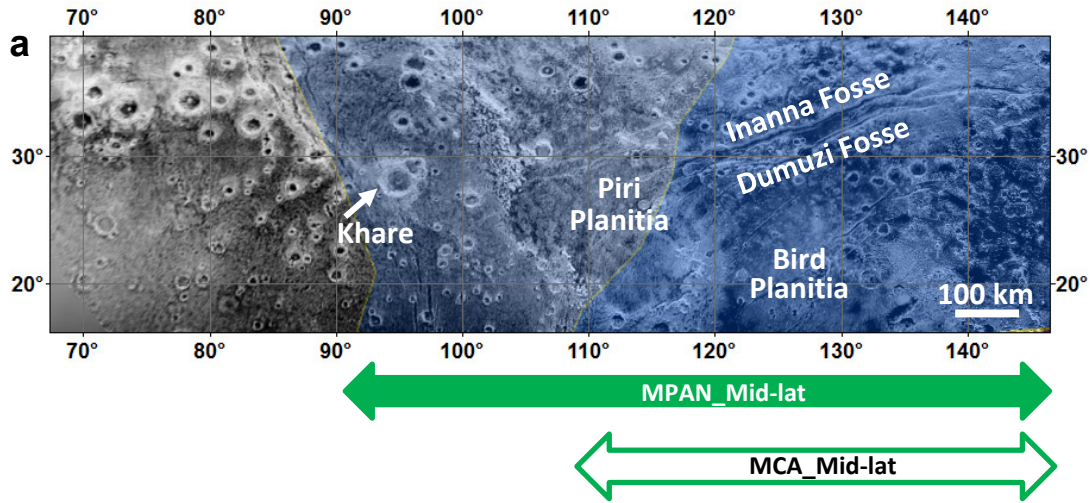

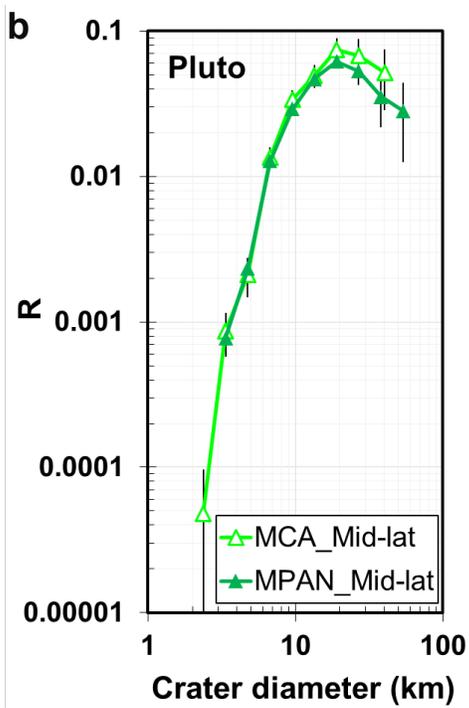

**Figure 3.3** (a) Simple cylindrical zoom of the mid-latitude region west of Sputnik Planitia showing varied crater densities and distinctive "bright halo craters". Background map from the 850 m px$^{-1}$ (PELR_P_LORRI) mosaic, with black outlines indicating the terrain covered by the higher resolution datasets whose crater measurements are shown in panel b. (b) SFDs from this region from the two different resolution datasets that cover different subsets of the area: PEMV_P_MVIC_LORRI_CA (315 m px$^{-1}$) and PEMV_P_MPAN1 (480 m px$^{-1}$).

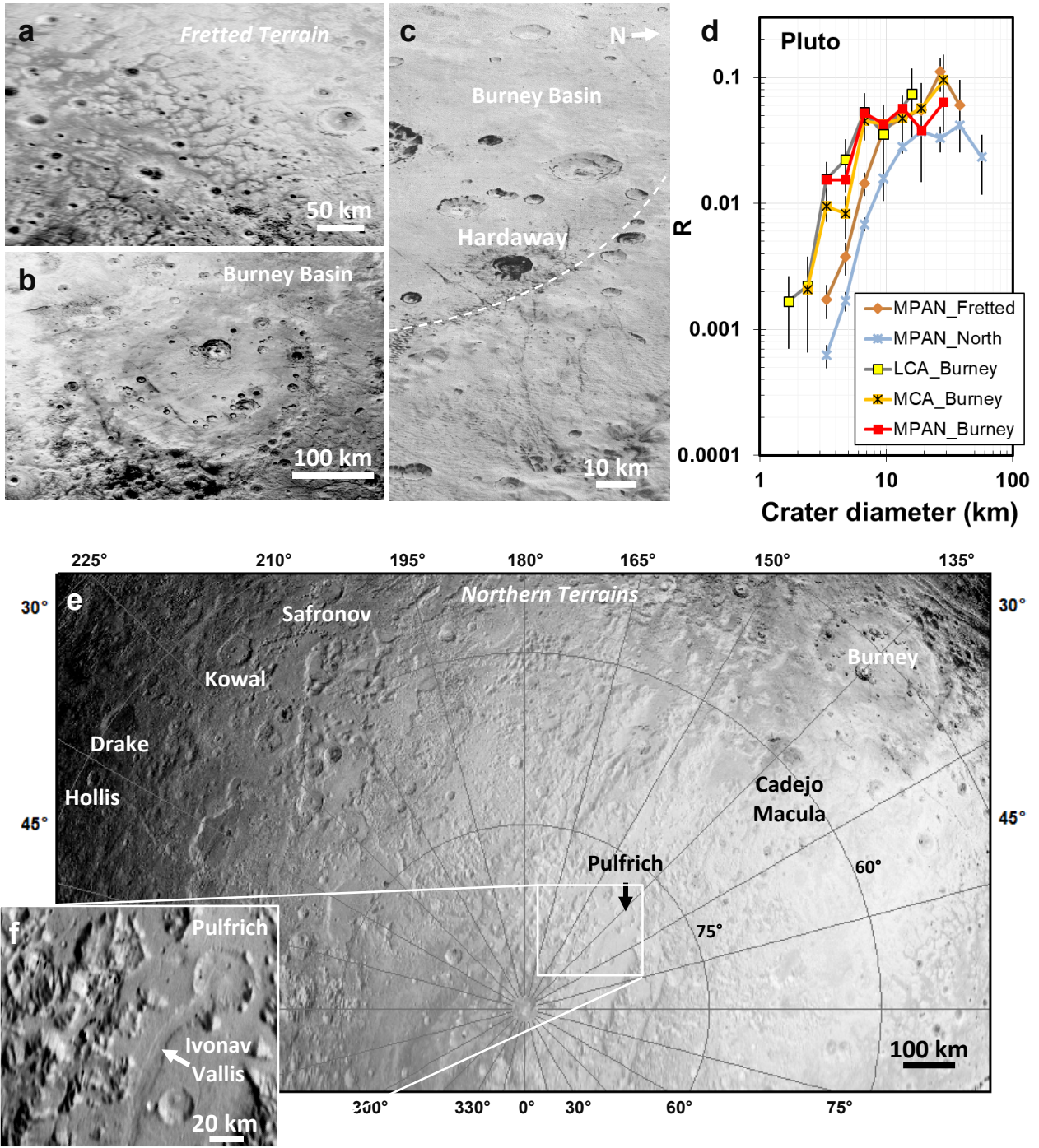

**Figure 3.4** (a) Fretted terrain and (b) Burney basin shown at 480 m px$^{-1}$ (PEMV_P_MPAN1). (c) A portion of the highest resolution strip at a pixel scale of 76 m px$^{-1}$ (PELR_P_MVIC_LORRI_CA) with part of Burney basin (dashed outline indicates approximate rim location) and superimposed younger craters, some with dark/bright layers visible in the walls, and one with distinct dark ejecta (Hardaway crater). (d) SFDs for the two northern regions (fretted terrain and the north polar area), and a comparison of the craters superimposed over Burney basin mapped at two different pixel scales. (e) North polar stereographic view of the varied northern terrain at 480 m px$^{-1}$ (PEMV_P_MPAN1), most of which is relatively lightly cratered and heavily mantled or resurfaced. (f) Several infilled craters exist in a large valley near Pluto's North pole.

*3.1.4. Lightly Cratered Terrains* Pluto has several regions with few-to-no identifiable craters. Eastern Tombaugh Regio (ETR) lies to the east of Sputnik Planitia and is dominated by rough, likely sublimation-driven, pitted terrain and perched, ponded glacial N$_2$ (*Moore et al.,* 2016; *Moore et al.,* 2017; also see chapter by Moore and Howard in this volume). One large potential crater in the center of ETR ($D$ = 30 km, labeled "A" in **Fig. 3.5a**), and several more along the edges are being eroded by the pit-forming process. On the few-to-10 km scale many pits form quasi-circular features. Most of these pits do not appear to be impact craters due to their lack of ejecta or distinct rims, and the fact that they share septa (dividing walls) with the surrounding pits. For the few circular features that appear to stand on their own, it is harder to determine if they are impact craters. Several deeper pit-like features (termed cavi) lie near Sputnik Planitia (labeled "B" and "C" in **Fig. 3.5a**). They are ~16 km and ~10 km in diameter, respectively, and ~3.5 km and ~2.3 km in average depth based on stereo topography (*Schenk et al.,* 2018c), giving them depth-to-diameter ratios of ~0.22. This is somewhat deep for craters of this size for icy surfaces, which are generally closer to a depth-to-diameter ratio of 0.1, but not implausibly so. The larger pit is not radially symmetrical in the height of the "rim" compared to the base. These features share the same craggy appearance as much of the rest of ETR, but they stand out because they are not filled with nitrogen ice, whereas much of the surrounding terrain is. Although the floors of these features are at or below the level of much of the surrounding smoother, nitrogen-ice-filled plains nearby, they do not appear to be filled with nitrogen ice themselves because they do not have flat floors. It is possible the surrounding craggy terrain is simply tall enough to keep N$_2$-rich ice from flowing into these depressions, whereas in other areas the N$_2$-rich ice is able to flow into the lows between the craggy peaks. Alternatively, the N$_2$-rich ice was removed from these pits by some mechanism (drainage or venting of pressurized gas are some possible mechanisms). These larger pits could be modified craters, or perhaps they formed through a completely different mechanism such as erosion, collapse, or venting from the subsurface. If they are modified craters, no ejecta or distinct rims are left for these features. Given the morphology of these deep pits is dissimilar in many respects from an impact crater, and is similar in many respects to the craggy terrain found elsewhere in ETR, we have not included them in the SFD displayed in **Fig. 5.2b**.

There are no distinct, unambiguous impact craters anywhere on Sputnik Planitia. The main body of bladed terrain (located on Tartarus Dorsa) is also devoid of obvious impacts, although one ambiguous ~32-km-diameter subcircular feature (labeled "D" in **Fig. 3.5e**) and several smaller ones exist closer to the edge of this terrain.

Few, if any, possible craters superpose the main mound of Wright Mons (*Singer et al.,* 2016). One possible 5.5-km-diameter crater sits in the ridged terrain near the rim of the Wright Mons central depression.  This one possible crater on the main mound has no obvious ejecta and is only quasi-circular with a non-continuous rim.  If it was formed as an impact crater, it would be degraded and/or deformed from its original form.  The wrinkly texture around the central depression is made of ridged structures that often look arcuate on small scales, and the ridges are emphasized in the roughly E-W direction by the oblique lighting (Wright Mons lies near the terminator).  Thus, it is not clear if this feature on Wright Mons is an impact crater, and it could alternatively be a collapse pit, ridge-structure, or cavus similar to others around the Mons.  There are a few small, more-circular possible craters in nearby terrains, but no other crater-like features are visible on the mound itself.  If this one 5.5-km-diamater feature is a crater, it is not fresh.  In addition to the characteristics described above, it has an accumulation of dark material in its interior.  It is possible, although unlikely, to have one 5.5-km-diameter crater form on Wright Mons, and for it to have time to degrade to its present state, while no other visible craters form. We present two cases for Wright Mons crater SFDs in **Fig. 5.2b**: one case where an upper limit is set by the lack of craters, and another case with this one 5.5-km-diameter feature as the only crater on Wright Mons.

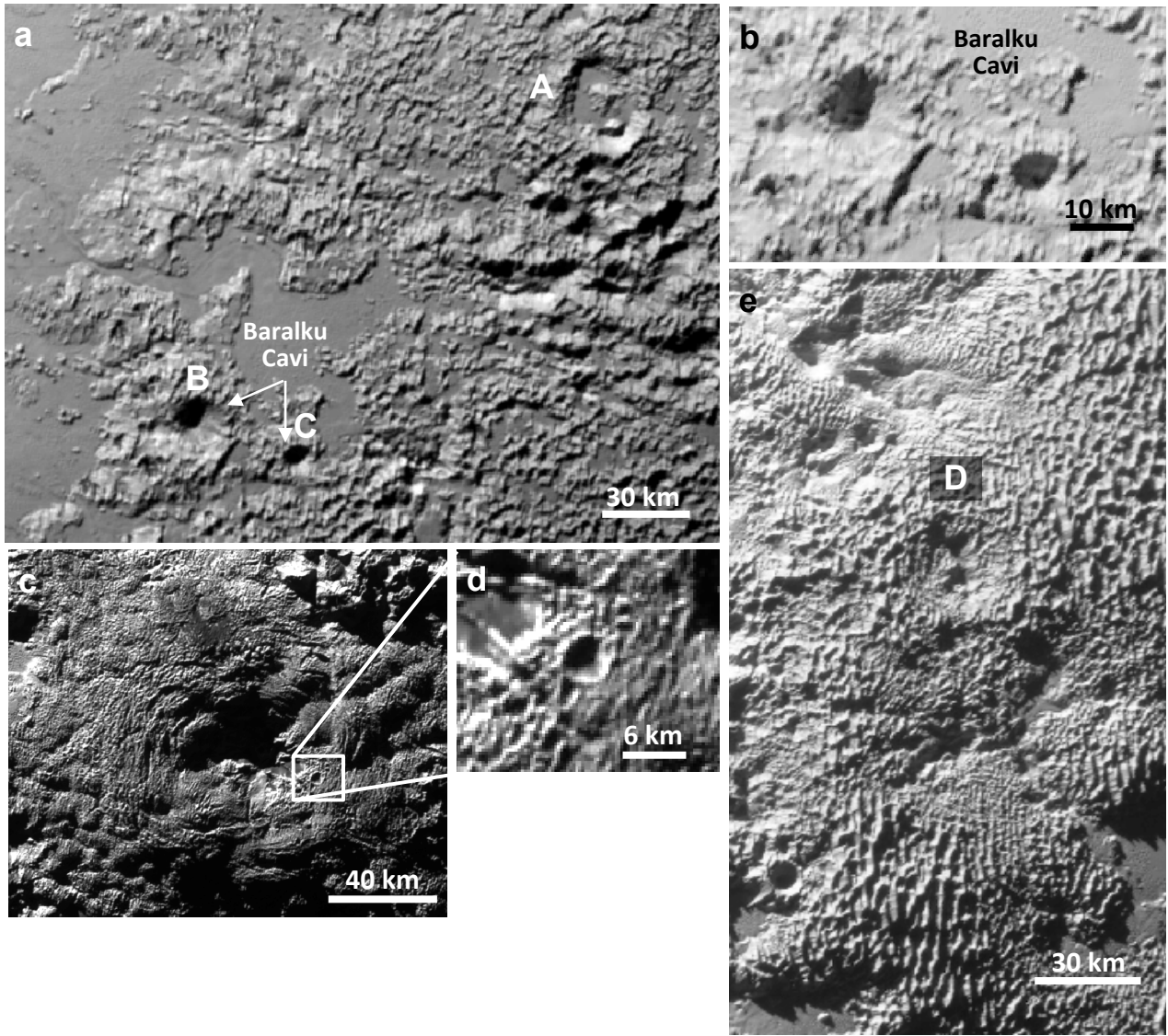

**Figure 3.5** Lower density cratered terrains include (a,b) eastern Tombaugh Regio, (c,d) Wright Mons, and (e) bladed terrain. A few of the larger ambiguous features are labeled with uppercase letters, including several pit-like structures informally named the Baralku Cavi described at length in Section 3.1.4. Panel (b) displays a different image stretch of Baralku Cavi to show the nature of the structures inside the shadowed regions. Image sources: all from PEMV_P_MVIC_LORRI_CA (315 m px$^{-1}$).

## 3.2. Charon Terrains

*3.2.1. Vulcan Planitia* Vulcan Planitia is a large plain that dominates the southern portion of Charon's encounter hemisphere (**Fig. 3.6**). In addition to craters, enigmatic "mountains in moats", narrow ridges and troughs, and also wider ropy structures decorate the smooth plains of Charon's Vulcan Planitia (**Fig. 3.6a**). In general, where craters occur, they seem to overprint all other features, and only a few examples of craters that may have been cut by tectonic activity exist (**Fig. 3.6d,e**).  Alternatively, it is possible the craters formed over a tectonic feature, which affected their final forms.  This may apply to the craters shown with an arrow in **Fig. 3.6e** where the craters are more comparable in size to the fracture they interact with.  Note that these two crater "halves" are unlikely to be part of a pit chain (formed through drainage of regolith into a fracture, rather than by impact) because they are offset from each other, are slightly different sizes, and have distinct rims.  Given there are very few examples of this type of feature, it is clear that most craters formed after tectonic episodes.

    None of the craters within Vulcan Planitia are partially filled, obviously embayed, or breached with smooth later flows (e.g., in the lunar mare style).  Just north of Vulcan Planitia on the eastern side (north of the large tectonic scarps defining the border) there are some partially filled craters; it is unclear if these flows were contemporaneous with the emplacement of Vulcan Planitia (and not as thick so the craters remain) or if they occurred somewhat later.  Some large craters within Vulcan Planitia do have hummocky floors, but they resemble landslide material seen in some lunar crater floors (see Fig. S4 in *Singer et al.*, 2019).  Some more lightly cratered areas exist across the plain.  These areas are conspicuous to the eye, but further statistical analysis is needed to determine if these low density areas could be stochastic or if they may indicate some later resurfacing after the majority of Vulcan Planitia was emplaced.

    Crater identification on Vulcan Planitia is relatively straightforward (*Robbins et al.,* 2017; *Singer et al.,* 2019) given the favorable lighting geometry (oblique sun) and the mostly smooth surface (**Fig. 3.6a**).  Small craters do exist, but are not abundant, and $D \lesssim 10$ km craters are deficient compared to a constant logarithmic slope distribution extrapolated from larger craters, similar to what is seen on Pluto.  The highest resolution strip on Charon (pixel scale of 155 m px$^{-1}$) yields similar results to the lower resolution datasets (*Robbins et al.,* 2017; *Singer et al.,* 2019).

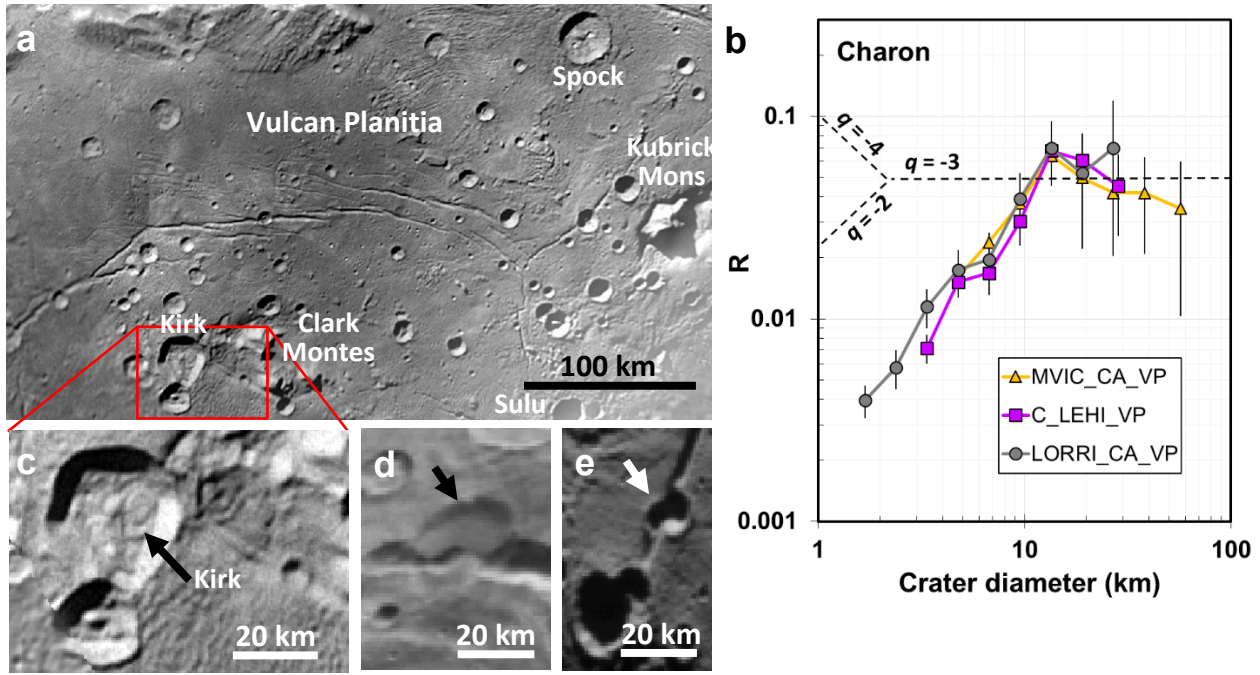

**Figure 3.6** (a) Charon's Vulcan Planitia, (b) SFDs for three different resolution datasets that cross Vulcan Planitia (also shown are dashed lines representing a differential slope of -4, -3, and -2 for reference), (c) close up of hummocky-floored craters, (d) one crater that may pre-date fracturing near 13°N/328°E, and (e) one or two craters that may have been affected by faulting on Vulcan Planitia (centered at 27°S/328°E), but see text for discussion. Image sources: panel a is from a mosaic of PELR_C_LORRI (865 m px$^{-1}$) and PELR_C_LEISA_HIRES (234 m px$^{-1}$; noted as C_LEHI in the legend), and c-e from PEMV_C_MVIC_LORRI_CA (622 m px$^{-1}$; noted as MVIC_CA in the legend). In the plot legend, LORRI_CA refers to crater data collected from the PELR_C_MVIC_LORRI_CA image set. Figure modified from *Singer et al.*, 2019.

*3.2.2. Oz Terra* The terrain north of Charon's smooth plains, collectively termed Oz Terra (**Fig. 3.7**; here we include the dark Mordor Macula north polar region in our SFDs), is a complex terrain with numerous scarps, valleys, large depressions, and a number of craters (*Beyer et al.,* 2017; *Robbins et al.,* 2019). The nearly overhead (low phase angle) lighting and lower signal-to-noise of the LORRI images of Charon's generally dark terrains make crater identification difficult in Oz Terra. Thus, we restricted the R-values as seen in **Fig. 5.3** to $D > 50$ km. Large structures were also examined with stereo topography. The large arcuate scarp in Mordor Macula (McCaffrey Dorsum indicated with small white arrows in **Fig. 3.7**) at first glance appears to enclose a basin, but topography reveals that the area interior to the scarp does not have a bowl-shaped, negative topographic expression (as would be expected for a basin; **Fig. 3.7c-d**), indicating the scarp may be instead a tectonic feature (*Beyer et al.,* 2017). A few smaller craters ($D < 30$ km) can be seen in Oz Terra, including distinctive craters that have dark inner-ejecta near the crater rim, and bright rays farther from the crater (some examples can be seen in **Fig.**

**2.2**). This ejecta pattern is consistent with layering in the near-subsurface of Charon that has been excavated by these craters (e.g., *Robbins et al.,* 2019). The material closer to the surface is ejected farther forming the bright rays, and deeper material (in these cases darker material) is ejected at lower speeds, landing closer to the crater. Several of these craters are associated with a stronger ammonia and water ice signatures in the New Horizons spectral data, implying this material is excavated from below (*Grundy et al.,* 2016; *Dalle Ore et al.,* 2018; chapter by Protopapa et al. in this volume).

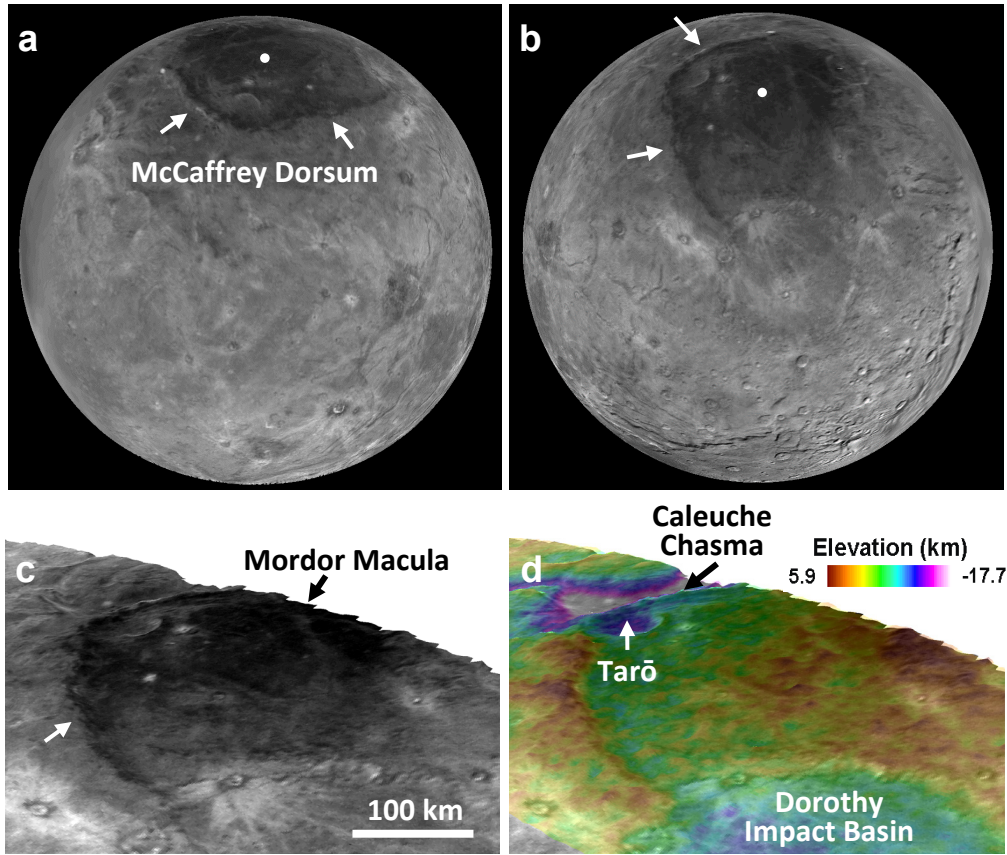

**Figure 3.7** The (a) western portion and (b) eastern portion of Oz Terra on Charon. White dot indicates Charon's northern or positive pole location. (c-d) Perspective views of Mordor Macula with heights projected by stereo topography (*Beyer et al.,* 2017; *Schenk et al.,* 2018b). Small white arrow in panels a–c indicates arcuate ridge discussed in Section 3.2.2. All panels from PELR_C_LORRI (865 m px$^{-1}$). The crater SFD for Oz Terra can be seen in **Fig. 5.3**.

## 3.3. Summary of Pluto and Charon Crater Size-Frequency Distributions and Relation to Impactor Populations

As can be seen above (and in **Figs. 5.2a and 5.3**), the crater SFDs on Pluto and Charon all show a break or "elbow" in the distributions, where larger craters have an average differential slope more similar to a power law exponent of $q = -3$ (*Robbins et al.,* 2017; *Singer et al.,* 2019), and craters smaller than ~10–15 km in diameter have a slope that is more shallow (closer to 0) than $q = -2$. A full table of slopes for each region is given in Table S2 of *Singer et al.*, (2019). For example, on Charon, the average slope for $D < 10$ km is $-1.7 \pm 0.2$. The Pluto regions show a larger variation in the slope below ~10 km diameter—consistent with the wider range of geologic processes occurring on Pluto. We believe this slope is representative of the impactor population, and not solely a product of differential erosion of smaller craters. This conclusion comes from several lines of evidence:

1. The crater SFDs on both Pluto and Charon show the same break location despite very different geologic histories between the two bodies (e.g., Charon does not have volatile ices on its surface or an atmosphere like Pluto does).
2. The same slope is seen for the eastern and western halves of Vulcan Planitia if it is split down the middle vertically (*Singer et al.*, 2019 Fig. S6).
3. The craters on Vulcan Planitia do not show any obvious signs of preferential erasure of small craters. Typically, a process that preferentially erases smaller craters would also leave some partially affected craters (e.g., intermediate-sized craters that are partially filled, embayed, or mantled), and those are not seen on Vulcan Planitia.
4. Although there are more geologic processes occurring on Pluto that could modify craters, there are only a few areas that show signs of preferential erasure of small craters (such as near the north pole; discussed in detail in *Singer et al.*, 2019). In most areas, craters are either completely erased or, similar to Charon, the more heavily cratered regions on Pluto do not all show intermediate-sized craters that are partially erased.
5. A similar shallow SFD slope for the equivalent size range of craters (taking into account scaling for different impactor velocities and surface gravities) is seen on some outer-planet satellite surfaces such as on Europa and Ganymede (e.g, *Zahnle et al.,* 2003; *Singer et al.,* 2019), and the Uranian satellites (*Kirchoff and Dones,* 2018).

*Singer et al.* (2019) examined the break location for Charon's Vulcan Planitia, because this is the surface that shows the least signs of crater modification between Pluto and Charon. They found the break to be at a crater diameter of ~13 km, which equates to an impactor size of $d$ ~1-2 km depending on which scaling parameters are used (see supplement of *Singer et al.* 2019).

A distinct break in slope around these sizes ($d$ ~1-2 km) is not seen for the asteroid belt, which in contrast retains a steeper differential slope (close to -3 or even steeper) at smaller sizes (see discussion in *Singer et al.*, 2019). This means the Kuiper belt has a large deficit of smaller objects compared to what would be expected if the SFD had a steeper slope (*Singer et al.,* 2019). In the smallest size bin measurable by the New Horizons crater data for Charon (the bin centered at 1.7 km in diameter in Fig. 3.6 for the LORRI_CA_VP distribution) there are only 30 craters found. There would be approximately an order of magnitude more craters expected (~300 total) if there were no "elbow" in the SFD.

New Horizons encountered the cold classical Kuiper belt object (486958) Arrokoth at 43 AU on January 1, 2019 (*Stern et al.,* 2019; also see chapter by Stern et al. in this volume). Arrokoth is a contact binary ~35 km across. By all indications the surface of Arrokoth is ancient, but it is not heavily cratered (*Singer et al.,* 2020a; *Spencer et al.,* 2020). This finding supports a lack of small objects in the Kuiper belt.

Nix was the best-imaged small satellite during New Horizons' 2015 flyby (*Weaver et al.,* 2016; also see chapter by Porter et al. in this volume). The Nix crater population is described in *Weaver et al.* (2016) and *Robbins et al.* (2017). Nix's surface has a relatively high spatial density of craters, equivalent to or higher than most heavily cratered terrains on Pluto or Charon, implying it is an old surface. Both the spatial density and number of craters are more uncertain on Nix than on Pluto or Charon because of the uncertainty in estimating the surface area on which craters were emplaced and the ambiguity of which circular, crater-like features are truly impact craters. Fitting a power-law to the data is also uncertain based on what data should or should not be included and an appropriate minimum diameter to which the data should be fit. Based on an examination of the most reliable features, fitting craters with diameters 2–6 km ($N$ = 15 features) from *Robbins et al.* (2017) yields a shallow differential slope of $q$ = –1.19±0.82. Because of the lower gravity on Nix, the same size impactor does not make the same size crater as it would on Pluto or Charon. Scaling Nix crates of $D$ = 2–6 km by a factor of 2.1 (as describe in *Weaver et al.,* 2016) would yield an equivalent diameter of craters of ~1–3 km on Pluto or Charon. While this uncertainty is large, the shallow slope is consistent with the shallow slope for $D$ < 10-15 km craters on Pluto and Charon.

The equivalent size-frequency distribution slope in the impactor population is, based on scaling theory, slightly shallower than that of the craters, and mildly dependent on the choice of material parameters chosen for the scaling from crater sizes to impactor sizes. All of the craters we can observe on Pluto or Charon should be in the gravity regime, and no matter which endmember materials are chosen (hard, non-porous ice or a more porous material like regolith/sand), only a small slope change is introduced. The full description of the scaling law derivation is in the supplement of *Singer et al.,* 2019 for Pluto and Charon. For the final power law scaling form of $d = a*D^b$ where $d$ is the impactor diameter, $D$ is the crater diameter, and $a$ is a scaling factor, $b$ comes out to be either 1.151 or 1.088, for a non-porous or porous target surface, respectively. The change in slope follows $q_{impactors} = q_{craters}/b$. For Charon's small crater slope of -1.7 ± 0.2, the equivalent differential impactor or Kuiper belt object (KBO) slope is approximately -1.5 (for $b$ = 1.151) or -1.6 (for $b$ = 1.088). This difference in slope between the impactor and crater distributions due to scaling is within the error bars of the slope itself but is mentioned here for completeness. The equivalent $H$-magnitude slope for a luminosity distribution of observed KBOs is described below, where we use the -1.5 value for the impactor slope for objects smaller than the "elbow" in subsequent analysis in this chapter.

This same conversion between crater and impactor SFD slopes applies for the larger craters, which display an average slope close to -3. However, there is a range of slopes for larger craters which also varies over the range of diameters available, as can be see in the SFD plots, and the error bars are large in some cases (*Singer et al.,* 2019, their Table S2). Thus, we continue to use the slopes as in *Greenstreet et al.* (2015) for modeling the cratering rates of larger craters (described more below), which is consistent with the crater data (also see discussion in supplement of *Singer et al.*, 2019).

## 4. IMPACTOR POPULATIONS AND RATES

### 4.1. Current Structure of the Kuiper Belt

The Kuiper belt is a reservoir of small bodies located beyond the orbit of Neptune, extending from 30 AU to ~ 1,000 AU with the majority of the classical and resonant sub-populations falling between roughly 30 AU and 85 AU. Pluto is the largest known Kuiper belt object (KBO). There are currently ~3,300 known KBOs with diameters $d \gtrsim 100$ m, but much of the population remains undiscovered due to their large distance from Earth. Thus, debiased surveys (those accounting for their observational biases) are currently the only available method for determining the intrinsic population of KBOs.

As more KBOs have been discovered, it has become clear that the Kuiper belt is divided into several dynamical sub-populations as described in *Gladman et al.* (2008), including objects in a mean-motion resonance (MMR) with Neptune, scattering objects that are actively experiencing deviations in semimajor axes due to current dynamical interactions with Neptune, and the classical or detached objects, which are neither resonant nor currently scattering off Neptune and are divided into several sub-components.

Pluto's orbit (semimajor axis $a \approx 39.5$ AU, eccentricity $e \approx 0.24$, and inclination $i \approx 17º$) puts it in the 3:2 MMR with Neptune (*Cohen and Hubbard,* 1965). Although Pluto's orbit (perihelion $q \approx 30.0$ AU and aphelion $Q \approx 49.0$ AU) crosses that of Neptune ($a \approx 30.1$ AU), the resonance protects Pluto from planetary close encounters with Neptune, allowing its orbit to remain stable for billion-year timescales.

### 4.2. Giant Planet Migration

The complex structure in the Kuiper belt, including many objects in MMRs with Neptune, the scattering and detached sub-populations on high-eccentricty (*e*), high-inclination (*i*) orbits, and the excited (moderate-*e* and moderate-*i*) hot component of the main classical belt, has numerous cosmogonic implications for solar system formation. The migration of the giant planets through a massive (10-100 Earth-mass) disk of planetesimals located from the giant planet region to the outer edge of the primordial Kuiper belt (*Fernandez and Ip,* 1984; *Malhotra,* 1993; *Hahn and Malhotra,* 2005) that is only halted when Neptune reaches the outer edge of the disk (*Gomes et al.,* 2005) is the leading explanation for much of the current structure of the Kuiper belt.

The Nice model is a currently heavily explored model of giant planet migration in the early solar system, which aims to reproduce the current orbital architecture of the giant planet system (*Tsiganis et al.,* 2005), the capture of the Jupiter (*Morbidelli et al.,* 2005) and Neptune (*Tsiganis et al.,* 2005) Trojan populations, and the precipitation of the late heavy bombardment of the terrestrial planets (*Gomes et al.,* 2005). The largest difficulty with the Nice model, however, is getting the massive planetesimal disk to remain for several hundred million years without either accreting into planets or collisionally grinding itself down into dust before the planets can disperse it into the current structure seen today. The 500 Myr delay in instability in the Nice model is thus now giving way (*Mann,* 2018) to a shorter phase with most of the rearrangement occurring in ≲ 1% the age of the solar system (see *Nesvorný,* 2018 for a recent review).

Once the migration of the giant planets ended, the sub-populations of the Kuiper belt have since naturally dynamically depleted at differing rates due to their differing orbital parameters over the past ≈ 4 Gyr. During this time period, it is assumed the orbital distribution of each population has not changed. A summary of the estimated sub-population decay rates from the literature can be found in *Greenstreet et al.* (2015). Due to the lack of knowledge about the detailed structure of the region beyond Neptune during the giant planet migration process, impact and cratering rates for the Pluto-system can only solidly rely on the orbital structure of the Kuiper belt currently known, which is believed to have been unchanged for the past ≈ 4 Gyr.

### 4.3. Size Distributions of Kuiper Belt Sub-Populations

There are large uncertainties in the Kuiper belt size distribution for objects with *g*-band absolute magnitude $H_g > 9.16$, where *g*-band has an effective wavelength of 463.9 nm and a width of 128.0 nm. Absolute magnitude (*H*) is defined to be the visual magnitude an observer would record if an object were placed both one astronomical unit (AU) away from the observer and 1 AU from the Sun at zero phase angle where the asteroid is fully illuminated. $H_g > 9.16$ corresponds to a diameter $d < 100$ km for a *g*-band albedo $p = 5\%$ using the equation $d \simeq 100$ km $\sqrt{(0.05/p)} * 10^{0.2(9.16-H\_g)}$. The differential number of objects *N* as a function of *H*-magnitude is defined by $dN \propto 10^{(\alpha H)}$, where $\alpha$ is the logarithmic "slope" (hereafter simply referred to as the slope) and maps to the differential distribution in object diameter *d*, $dN \propto d^{(q)}$, by $-q = 5\alpha + 1$.

The Kuiper belt size distribution for $H_g \gtrsim 8\text{-}9$ has been absolutely calibrated by the Canada France Ecliptic Plane Survey (CFEPS) (*Petit et al.,* 2011; *Gladman et al.,* 2012) and is well represented by a single logarithmic "slope" $\alpha$ for all sub-populations with the exception of the hot and cold components of the main classical belt, which appear to have different values of $\alpha$ (*Bernstein et al.,* 2004; *Bernstein et al.,* 2006; *Petit et al.,* 2011; *Adams et al.,* 2014; *Fraser et al.,* 2014). Extending to KBOs smaller than $H_g = 9$, it is clear a single power law does not fit the observations and a break in the differential size distribution at this $H_g$-magnitude is needed, which is explored in great detail in: *Jewitt et al.* (1998), *Gladman et al.* (2001), *Bernstein et al.* (2004), *Fraser and Kavelaars* (2008), *Fuentes and Holman* (2008), *Shankman et al.* (2013), *Adams et al.* (2014), and *Fraser et al.* (2014). Also see chapter by *Parker* in this volume for additional information on observation surveys of KBOs.

Due to the uncertainty in the Kuiper belt size distribution for $H_g \lesssim 9$, one must assume a size distribution when computing impact and cratering rates onto Pluto and Charon, which provides the largest source of uncertainty for the resulting rates. In the past, groups have modeled a variety of assumed impactor size distribution slopes and compared the resulting crater SFDs, each producing slightly different predictions. The uncertainties in the impactor size distribution are manifested in calculated terrain ages that use the various predictions for absolute calibration. Terrain age estimates for Pluto and Charon will be discussed in Section 5.

**4.4. Computing Collision Probability**

Soon after the Kuiper belt was discovered, impact rates onto Pluto and Charon began emerging in the literature. *Weissman and Stern* (1994), *Durda and Stern* (2000), and *Zahnle et al.* (2003) were some of the first to produce estimates. Their methods consisted of particle-in-a-box calculations or approximations of the KBO number density for those that intersect Pluto's orbit at an average impact speed. *de Elia et al.* (2010) computed the impact flux of Plutinos (KBOs other than Pluto located in the 3:2 MMR with Neptune) onto Pluto assuming typical impact speeds of 1.9 km s$^{-1}$ and used this to calculate cratering rates onto Pluto from the Plutinos alone.

*Dell`Oro et al. (2013)* computed impact probabilities of the individual KBO sub-populations from the CFEPS L7 model (*Petit et al.,* 2011; *Gladman et al.,* 2012) through collisional evolution but did not extend their analysis to the cratering rate onto Pluto. The extension of the *Dell'Oro et al.* (2013) cratering rate of Plutinos onto Pluto assuming a mean impact velocity was performed by *Bierhaus and Dones* (2015), who also included the production of secondary and sesquinary craters onto the surfaces of Pluto and Charon into their analysis (see Section 4.6 for more on secondary and sesquinary craters). Around the same time, *Greenstreet et al.* (2015, 2016) computed the impact and cratering rates onto Pluto and Charon using a similar combination of KBO sub-populations to those used in *Bierhaus and Dones* (2015) but with a different assumed KBO size distribution for objects smaller than telescopic surveys have observed ($d \lesssim 100$ km), taking into consideration the unique dynamics of Pluto's orbit in the Kuiper belt. *Greenstreet et al.* (2015) computed the impact probability onto Pluto and Charon by modifying a version of the Öpik collision probability code that implements the method described in *Wetherill* (1967) and is based on *Dones et al.* (1999).

Most of the impact and cratering rates onto Pluto and Charon in the literature assume an average impact speed. *Zahnle et al.* (2003) and *Dell'Oro et al.* (2013) quote typical impact speeds from KBOs onto Pluto (or the Plutinos) to be approximately 2 km s$^{-1}$. *Bierhaus and Dones* (2015) adopt this average impact speed for their analysis. *Greenstreet et al.* (2015) modified their analysis to produce a spectrum of impact speeds for each Kuiper belt sub-population. Compared to other cratered bodies in the outer solar system that have been studied to date, Pluto uniquely sits within the Kuiper belt. Due to the detailed orbital architecture of the various sub-populations within the belt, each sub-population has a different impact probability and thus impact speed onto Pluto. This is emphasized by the complex dynamics Pluto's orbit experiences over time (see *Greenstreet et al.,* 2015 for a more detailed discussion). **Figure 4.1** shows the impact velocity spectrum onto Pluto from *Greenstreet et al.* (2015). They find a mean impact speed of 2 km s$^{-1}$ for the combined KBO population, but show that the various sub-populations produce a wide spread in impact speeds from Pluto's escape speed at 1.2 km s$^{-1}$ out to a tail at 5 km s$^{-1}$. Before turning their impact speeds into averages, *Dell'Oro et al.* (2013)

computed impact speed distributions for the collisional evolution of the KBO sub-populations. Their impact speed distribution for the Plutinos onto each other independently reproduces the same main trends found by *Greenstreet et al.* (2015) for the Plutinos impacting Pluto.

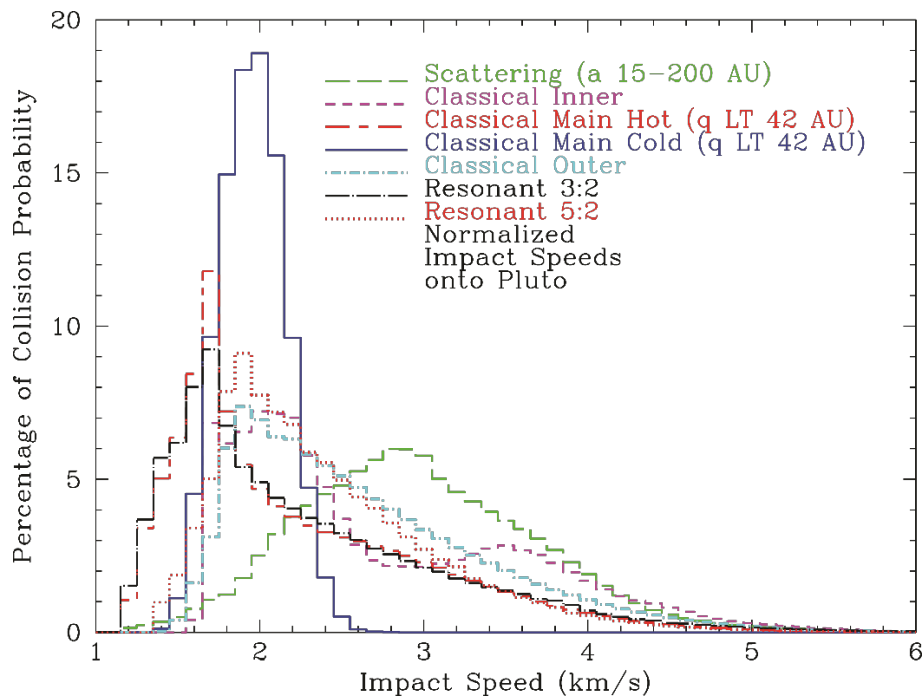

**Figure 4.1** Impact speed distributions onto Pluto with data from *Greenstreet et al.* (2015). Escape speed from Pluto is 1.2 km s$^{-1}$. Each sub-population's distribution is separately normalized.

### 4.5. Impact Rates Onto Pluto and Charon

**Table 4.1** lists the computed impact rates for Pluto and Charon described below. *Weissman and Stern* (1994) provided estimates of $d > 2.4$ km comets impacting Pluto and Charon at present rates. For KBOs with $d > 2$ km, *Durda and Stern* (2000) estimated impacts occur on Pluto and Charon on shorter timescales. In addition, they computed the impact rate for KBOs with $d > 100$ km on Pluto.

*Zahnle et al.* (2003) computed impact rates onto Pluto and Charon for $d > 1.5$ km comets. For $d > 100$ km impactors, *Zahnle et al.* (2003) provided an impact rate scaled from the calculations of *Nesvorný et al.* (2000) for plutino impacts onto Pluto and an Öpik-style collision probability estimate from W. Bottke (personal communication). *Zahnle et al. (2003)* also reported an impact rate on Charon that is 16% that on Pluto for $d > 100$ km impactors. *de Elía et al.* (2010) calculated impact rates for Pluto and Charon for $d > 1$ km plutinos.

*Bierhaus and Dones* (2015) estimated impact rates for Pluto and Charon using the estimates from *Dell'Oro et al.* (2013) for plutinos with $d > 1$ km colliding with each other. *Greenstreet et al.* (2015) provided an impact rate for $d > 100$ km impactors onto Pluto from a similar combination of Kuiper belt sub-populations to those used in *Bierhaus and Dones* (2015). In addition, *Greenstreet et al.* (2015) reported an impact rate 19% that on Pluto for $d > 100$ km impactors on Charon.

| Reference | Impactor Diameter Range (km) | Pluto Impact Rate ($\times 10^{-8}$/yr) | Charon Impact Rate ($\times 10^{-8}$/yr) |
|:---:|:---:|:---:|:---:|
| Weissman & Stern (1994) | > 2.4 | 53 | 10 |
| Durda & Stern (2000) | > 2 | 260 | 31 |
| Zahnle et al. (2003) | > 1.5 | 100 – 250 | 16 – 42 |
| de Elia et al. (2010) | > 1 | 36 – 160 | 10 – 44 |
| Bierhaus & Dones (2015) | > 1 | 46 | 10 |
| Durda & Stern (2000) | > 100 | 0.012 | --------- |
| Zahnle et al. (2003) | > 100 | 0.0023 | 0.00037 |
| Greenstreet et al. (2015) | > 100 | 0.0048 | 0.00092 |

**Table 4.1** Pluto and Charon impact rates for a range of impactor diameters from the literature. See text for more details on each study.

Both *Bierhaus and Dones* (2015) and *Greenstreet et al.* (2015) broke down their impact rates by Kuiper belt sub-population and noted the importance of each to the total impact rate on Pluto. Both papers found that the Cold Classical KBOs dominate the Pluto impact flux, but by vastly differing amounts (≈84% (*Bierhaus and Dones*, 2015) versus ≈37% (*Greenstreet et al.*, 2015) for $d \geq 10$ km) given their different assumed KBO size distributions.

Although these published estimates span a time period of more than two decades, they are, surprisingly, in agreement to within a factor of 3 for both Pluto and Charon. This means terrain ages calculated from these estimates agree to within a factor of 3-4. Although this disagreement is relatively small, this could mean a range of 1-4 Gyr in the age estimate of a given surface area on either body, for example. Thus, the assumptions and considerations that went into the predictions should be considered when used for interpretation of the observational data.

## 4.6. Cratering Rates Onto Pluto and Charon

To convert impact rates to cratering rates, a crater scaling law is needed. The properties of the target and impactor both play a role in the size of crater produced for a given size impactor (e.g., *Holsapple,* 1993). Crater scaling laws are developed from a combination of empirical data and physics principles (*Housen and Holsapple,* 2011). The empirical data includes both laboratory work and measurements of observed craters and ejecta, both from natural impacts and man-made explosion cratering. The crater scaling law used in *Greenstreet et al.* (2015) from *Zahnle et al.* (2003) was developed for solid, non-porous geological materials (relevant to the icy surfaces of Pluto and Charon). As an example, the scaling from *Zahnle et al.* (2003) produces roughly 3% larger simple crater diameters than the scaling law from *Housen and Holsapple* (2011), which was used in *Bierhaus and Dones* (2015). However, as discussed in Section 3.3 above, the difference in slope between the impactor and crater distributions due to different scalings is small and within the error bars of the slope itself.

As mentioned above, *Bierhaus and Dones* (2015) included the production of secondary and sesquinary craters in their computation of the cratering rates onto Pluto and Charon. Secondary craters are produced by debris fragments ejected during the formation of the primary crater that impact outside of the primary crater at high enough velocity to form their own crater. Sesquinary craters are made when ejected material is travelling fast enough to escape the source body and subsequently re-impact that body or another object at a later time. They concluded that sesquinaries are not expected to be an important component of the overall Pluto or Charon crater SFDs and secondaries should be visible as a steeper branch of the crater SFD at diameters less than a few km on Pluto and are likely not to be visible in Charon's observed crater SFD.

To produce a crater size-frequency distribution, one must integrate over the impactor size distribution to convert impactor diameter to crater diameter. The resulting crater SFD then depends on the assumed KBO size distribution. If an average impact speed is used, the crater scaling laws give a single crater diameter for each impactor diameter. In the case of an impact velocity spectrum, the impactor-to-crater diameter conversion is no longer a one-to-one relationship. A given size impactor produces craters with a variety of diameters when travelling at a range of impact speeds. To account for this, one must integrate down the impact speed distribution as well as the impactor size distribution in the conversion of impact rates to cratering rates onto Pluto and Charon (*Greenstreet et al.,* 2015).

*Bierhaus and Dones* (2015) suggested two possible crater SFD slopes: (i) slope of $q \sim -2$ ($\alpha = 0.2$) from the young terrains of the Galilean and saturnian satellites that indicate a shallow distribution; (ii) slope of $q \sim -3$ ($\alpha = 0.4$) from observational data of KBOs extended from $d \approx 100$ km down to smaller sizes. *Greenstreet et al.* (2015) used several KBO size distribution models to illustrate the uncertainty in the crater SFD slope. Their models included a power law with a "knee" at $H_g=9.0$ that has a sudden slope change from $\alpha_{bright} = 0.8$ ($q = -5$) to $\alpha_{faint} = 0.4$ ($q = -3$), a power law with a sudden drop in the differential number of objects (i.e., a "divot") by a factor of 6 with the same $\alpha_{bright}$ and $\alpha_{faint}$ as the "knee" distribution, and the "wavy" size distributions from *Minton et al.* (2012) and *Schlichting et al.* (2013). The resulting crater SFDs for Pluto and Charon are shown in **Fig. 4.2** in the form of a relative crater frequency R-plot, which is normalized to a differential $D^{-3}$ distribution (as described in Section 3). The predicted crater SFD from *Zahnle et al.* (2003) based on young surfaces on Europa and Ganymede is also shown in **Fig. 4.2**. Note that the two slopes in the *Zahnle et al.* (2003) broken power law are the same as the two separate slopes used in *Bierhaus and Dones* (2015). It should be also noted that because the escape speeds (Pluto: $v_{esc}$ = 1.2 km s$^{-1}$, Charon: $v_{esc}$ = 0.675 km s$^{-1}$), and gravitational accelerations (Pluto: $g$ = 64 cm s$^{-2}$, Charon: $g$ = 26 cm s$^{-2}$) are different for Pluto and Charon and both are factors in the crater scaling laws, a given impactor will produce a slightly larger crater on Charon than on Pluto, shifting the crater SFDs. A discussion of the uncertainties in the crater SFDs can be found at the end of Section 4.7.

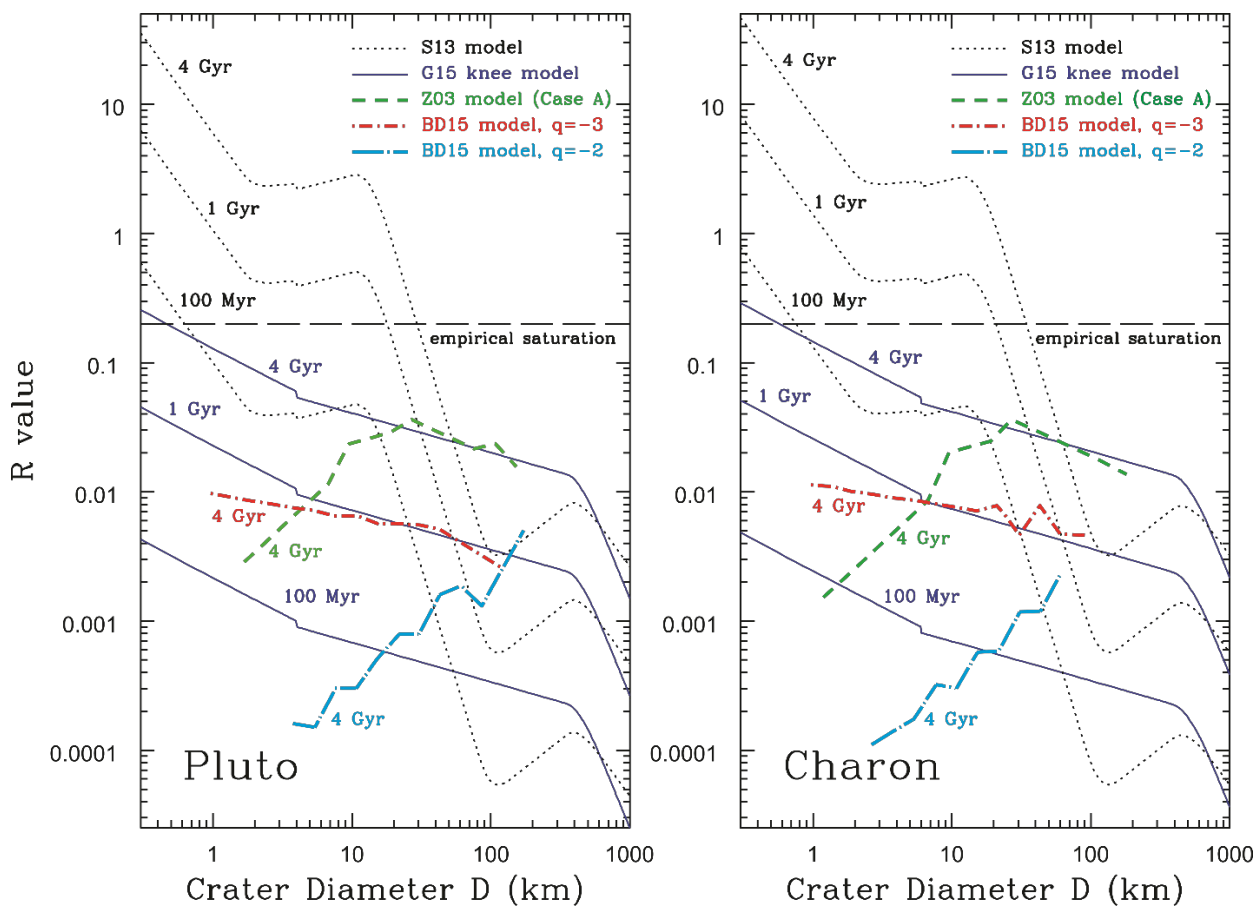

**Figure 4.2** Relative crater frequency R-plot including KBO size-frequency distribution predictions from *Zahnle et al.* (2003) (Z03), *Schlichting et al.* (2013) (S13), *Bierhaus and Dones* (2015) (BD15), and *Greenstreet et al.* (2015, 2016) (G15) for Pluto (left) and Charon (right). Predicted crater SFDs from the G15 and S13 models show three different age surfaces. The empirical crater saturation density calculated in *Greenstreet et al.* (2015) from *Melosh* (1989) is shown for reference, where crater densities typically do not increase due to the erasure of previous craters from newly forming ones.

## 4.7. Predictions and Implications for Crater SFDs in the Pluto-system

In anticipation of the New Horizons flyby of Pluto and Charon in 2015, *Bierhaus and Dones* (2015) and *Greenstreet et al.* (2015, 2016) made predictions about the cratered surfaces of both bodies. They reported that the expected visible range of craters on both Pluto and Charon is *D* ~1-100 km given the uncertainty in the impactor size distribution. *Bierhaus and Dones* (2015) point out that if the crater SFDs are like those of the Galilean and saturnian satellites, young unsaturated terrains should show a shallow differential slope of $q \sim -2$ ($\alpha \sim 0.2$). This would imply there is no process present that erodes small KBOs as they migrate inward from the region beyond Neptune to the giant planet region and also that small KBOs are deficient relative to an extrapolation of larger KBOs visible to telescopic surveys down to smaller sizes. This last point would have important implications for constraints on the evolution and possible formation mechanism of the KBO sub-populations. However, if the $q \sim -3$ ($\alpha \sim 0.4$) slope from large ($d \sim$ 100 km) KBOs extends down to small sizes ($d < 1$ km), then there will be a steep ($q \gtrsim -3$) crater SFD observed across all crater sizes (*Bierhaus and Dones*, 2015; *Greenstreet et al.*, 2015). This would mean some process could exist which destroys small KBOs as they move to smaller heliocentric distances near Jupiter and Saturn making them under-abundant at these distances (*Bierhaus and Dones*, 2015).

Both *Bierhaus and Dones* (2015) and *Greenstreet et al.* (2015) argue that even in 4 Gyr of bombardment, Pluto and Charon are not expected to be saturated for *D* > few km craters, making it possible to at least relatively (or model-dependently) date their surfaces. *Bierhaus and Dones* (2015) add that if few secondary craters are visible, surfaces may not be saturated at any crater diameters observable by New Horizons. Only one of the impactor size distributions would predict saturation of Pluto's surface: the *Schlichting et al.* (2013) size distribution as explored in *Greenstreet et al.* (2015), which if correct would mean Pluto's surface would saturate for D $\lesssim$ 15 km in only 1 billion years of bombardment. However, as shown in the observed crater SFDs for Pluto and Charon provided in Section 3, a deficit of craters is found for this size range (*D* $\lesssim$ 15 km) ruling out the *Schlichting et* al. (2013) size distribution model for impactors of this size.

There are various sources of uncertainty in the predicted cratering rates: (i) the choice of the various impactor size distributions assumed (which translate into the slopes of the predicted crater SFDs), (ii) the impactor population estimates used for normalizing the size distributions (manifested in the vertical scaling of the crater SFDs), and (iii) the different crater scaling laws, which shift the crater SFDs to slightly different crater diameters. The compilation of these factors leads to a roughly factor of two uncertainty in predictions of the crater SFDs on Pluto and Charon.

The largest uncertainty in the cratering rates and thus terrain ages computed for the Pluto system is the extrapolation of the KBO size distribution to diameters d $\lesssim$ 100 km (corresponding to a crater diameter of D $\lesssim$ 400 km on Pluto; *Bierhaus and Dones*, 2015; *Greenstreet et al.*, 2015). New Horizons observations probe impactor diameters below the current observational limits through the craters observed on both Pluto and Charon, providing the first observational opportunity to determine relative slopes and population estimates of the d ≲ 100 km impactor population. As discussed in Section 3.3, the information from New Horizons revealed an additional break in the impactor/crater SFD at sizes smaller than can be seen with Earth- or space-based telescopes. We have added this new information into the terrain age predictions discussed in the next section.

## 5. TERRAIN AGES

Here we combine the measured crater size-frequency distributions for Pluto and Charon (Section 3) and predictions for the crater spatial density of a given age terrain (Section 4) to produce estimates of the crater retention age for different terrains. The crater retention age refers to how old the observed surface (and near-subsurface) is with respect to formation and removal of craters. A surface with no identifiable craters (down to the limits of our image resolution) means that all craters have been removed by resurfacing processes that may have been ongoing for a long period of time, or may have been removed in one or more discrete events in the relatively recent past. These surface age estimates use the starting place that all craters on Pluto and Charon were formed by single heliocentric impactors and not from secondary or sesquinary debris, or from binary primary impactors. As discussed below in Section 5.1 (and in *Singer et al.* 2019), we see no strong signs of secondary craters or circumbinary debris.

Binary objects are observed in the Kuiper belt by telescopic surveys, but observational constraints limit our knowledge of their occurrence rate (see discussion in the chapter by *Parker* in this volume). Statistical analysis of whether a binary impactor population can be seen in the crater populations on Pluto and Charon is discussed in the chapter by *Parker* in this volume. A large binary fraction in the impactor population would affect the resulting crater densities on Pluto and Charon and thus the estimated terrain ages. If, for an extreme case, 50% of the impactor population is a near-equal-size binary, this binary fraction could shift observed crater densities to higher values by as much as 50% more than crater densities produced by only single impactors. Not all binaries would necessarily produce distinct craters, because that would depend on the orientation of the binary as it impacted and the separation between the two objects, but the example here gives the maximum effect. This would result in the corresponding terrain ages shifting to 50% shorter than that for an impactor population consisting of only single impactors due to the decreased amount of time required to reach the same spatial density of craters. In simple terms, the equation for including the influence of binaries is: (single impactor terrain age estimates) / (1 + binary fraction) = (binary impactor terrain age estimates). For example, a terrain age estimate of 4 Gyr for a population of only single impactors would correspond to a terrain age estimate of 4 / 1.5 = 2.7 Gyr for a population consisting of 50% binaries. This simple equation can easily be scaled for any binary fraction within the impacting population.

In the more realistic case that all binaries do not consist of objects of the same size, the above scaling is an upper limit of the affect of binary impactors on terrain age estimates. If binary pairs are not the same size, the craters each binary produce will not be the same size, decreasing the rate at which craters of a given size are accumulated on the surface of either Pluto or Charon. This results in a smaller shift in crater densities compared to an impactor population with only single impactors (i.e., by <50% for a 50% binary impactor fraction) and thus a smaller shift in the corresponding terrain age estimates. The size distribution of the binaries would be required to determine the quantitative shift in the terrain age estimates for an impactor population consisting of any fraction of binaries compared to that of a single-impactor-only impacting population.

We emphasize that all ages reported here are rough, order-of-magnitude estimates, and the uncertainties in these results depend on several factors. There is uncertainty in the predicted cratering rate (described above in Section 4.6) of roughly a factor of 2 (*Greenstreet et al.*, 2015). The impactor flux and cratering rate models will be continually refined as more observations of the Kuiper belt are made. Here we do not pick a specific crater size to pin our estimates to (as is done for some other bodies in the solar system) because that is not practical with our datasets. Thus, the estimates below are either given as upper or lower limits, or as rough best answers given the limitations of the data and models.

Determining the age of surfaces with few or no craters is an additional challenge, but has been discussed for other bodies (e.g., *Michael et al.,* 2016). It is clear that the large areas of Pluto's encounter hemisphere that are devoid of craters are younger than the cratered regions, and some have additional age constraints based on modeling of their ongoing activity, such as Sputnik Planitia (*McKinnon et al.,* 2016). Here we discuss age estimates for the younger terrains based on the *Greenstreet et al.* (2015; 2016) "knee" model modified with an additional break in slope for craters $D < 13$ km (we will call this slope break the "elbow") (see **Figs. 5.2 and 5.3** and *Singer et al.*, 2019). We use the average slope of craters on Charon's Vulcan Planitia, $q = -1.7$, as a representative slope for this additional piece of the SFD power law to modify the result of *Greenstreet et al.* (2015; 2016). We use a modified version of the *Greenstreet et al.* (2015; 2016) "knee" model as our reference model here, because it is the most recent cratering rate prediction based on telescopic observations of the Kuiper belt subpopulations, and because this extrapolation of the impactor size distribution most closely matches the constraints found by New Horizons (e.g., the actual crater distributions found). The *Zahnle et al.* (2003) base prediction is fairly similar to the *Greenstreet et al.* (2015; 2016) as can be seen in **Fig. 4.2,** and thus would produce similar results**.** All of these models, and the terrain ages predicted by them, are subject to future revisions.

Each case is described below, but for surfaces with no craters, we give an *upper limit* constraint on the crater retention age based on the smallest crater that could be seen with the available image resolution and the area of that unit. This age constraint is an upper limit because there may, or may not be, smaller craters on these terrains that we cannot see because our image resolution is too course. If, in reality, no smaller craters exist on these surfaces, the derived age constraint would be even younger than what we report here. This method is conceptually similar to that in *Michael et al.* (2016) for terrains with no craters, but we do not have a full chronology function and so our approach is somewhat simpler. Several of the terrains described below have similar upper age constraints because the image resolution is similar over these terrains (and they are also somewhat similar in areal extent). This does not necessarily mean that they were all resurfaced at the same time, rather that they were resurfaced sometime between the present and the upper age constraint. In a few cases superposition relationships or other information can be used for additional terrain timing relationships (also described in Section 3 and in the chapter by *White et al.* in this volume).

**5.1. A Note about Secondary Craters and Crater Ejecta**

Pluto has a striking lack of obvious secondary craters, or even noticeable crater ejecta (except for a handful of cases, see **Fig. 2.1**), at all the available image resolutions. Obvious secondary craters are defined here as craters in clusters or chains, or craters/clusters with radial indicators such as v-shaped ejecta or elongation/asymmetry that point back to a primary crater (some examples are given in **Fig. 5.1**). Not all secondary craters form with these morphological features, and some of these features can be modified or lost over time, but they are general indicators of secondary craters. Secondary craters also tend to have steep SFD slopes (as steep as $q = -6$), specific spatial patterns (e.g., radial distribution around a primary crater, a chain or cluster that points to a larger primary crater), and a specific size relationship to the primary crater—the very largest secondaries are generally not more than ~5-8% the size of the primary (e.g., *Melosh,* 1989; *Singer et al.,* 2013). There may be secondary craters below our resolution limits or non-obvious secondary craters. Both obvious and non-obvious secondary craters would tend to steepen the SFD at small crater sizes, as is seen on Europa (*Bierhaus et al.,* 2009), but we observe no steepening of the SFD. Although pits exist on Pluto at many scales, and pit chains occur over likely fractures, none of these pits appear to be secondary craters. On Charon, ejecta deposits are more visible indicating a slower process of erosion than on Pluto. However, there are no features resembling traditional secondary craters on the smooth plains of Vulcan Planitia. There is only one possible feature seen on Oz Terra that could represent a crater chain: a narrow catena-like chain located at ~30°N/10°E (**Fig. 5.1d**). Lighting geometry, image resolution, and terrain characteristics (such as large albedo variations or local geology) can also affect how all features, including secondary craters appear. For example, some image conditions can make the smaller, more subtle feature such as v-shaped ejecta more difficult to see. However, the resolution and lighting geometry varies greatly across both Pluto and Charon, and at the sizes of the mapped craters, few-to-no secondary-crater-like features are seen. Because of the reasons listed here, we believe secondary craters are not a large contribution to the crater data presented here.

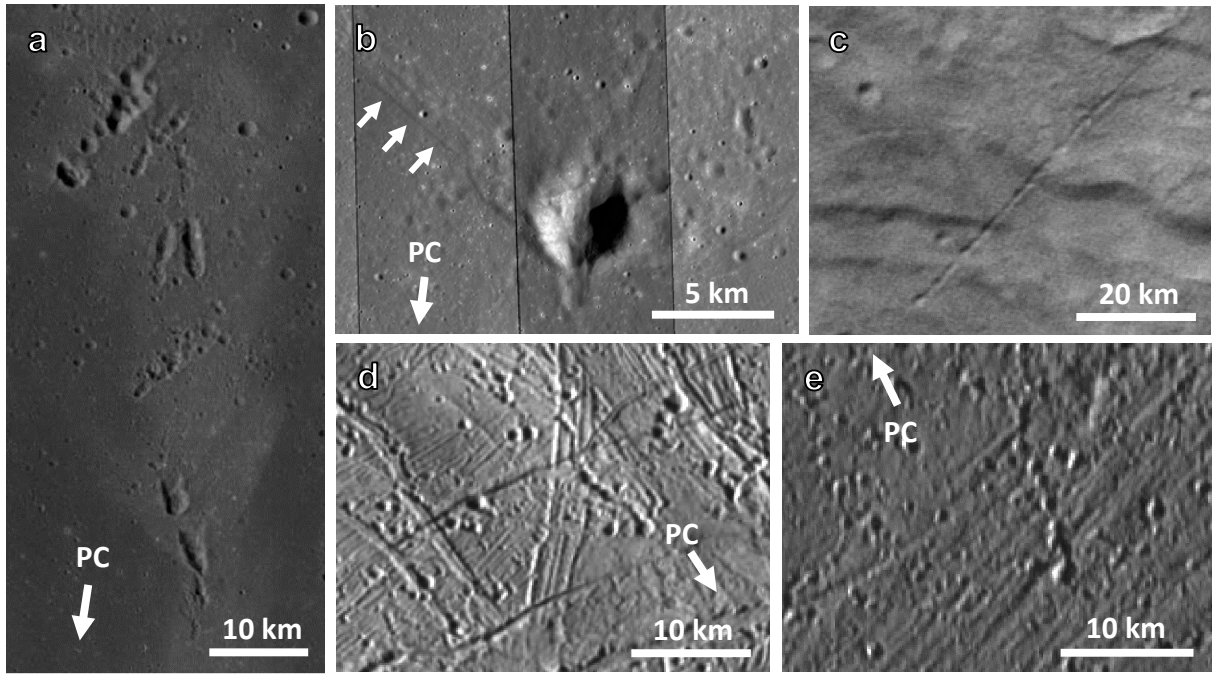

**Figure 5.1** Example secondary craters from other bodies in the solar system. Very few features on Pluto or Charon appear to be secondary craters, but they may exist below the resolution limit of our images. Note in addition to being clusters/chains, these features are found either radial to or nearby a primary crater and are the expected sizes for secondary craters around their respective primary. The arrows noted with "PC" indicates the direction to the primary crater. (a) Classic examples of secondary crater clustering and morphology around the lunar crater Copernicus (93 km in diameter; more details in *Singer et al.,* 2020a) as seen in the Lunar Reconnaissance Orbiter Camera (LROC) Wide Angle Camera mosaic at 100 m px$^{-1}$. (b) Close up of a relatively-unmodified lunar secondary crater (also thought to be from Copernicus) where v-shaped ejecta can be seen (3 arrows in a row give an example). Additionally, the rim of the crater in the downrange direction is less well-formed, which is a common feature of secondary craters that are found relatively close to their primary. This image is from a LROC Narrow Angle Camera images shown at ~1.3 m px$^{-1}$ (modified from *Singer et al.,* 2020b). (c) Charon crater chain on Oz Terra centered at 10.2°E, 30.8°N (from PEMV_C_MVIC_LORRI_CA, 622 m px$^{-1}$). (d) Secondary craters around the Tyre impact basin on Jupiter's moon Europa shown at ~210 m px$^{-1}$ (*Bierhaus et al.,* 2009; *Singer et al.,* 2013) (e) Secondary craters around the Achelous crater on Ganymede shown at ~180 m px$^{-1}$ (*Singer et al.,* 2013).

Here we also consider the size of secondary craters that might be expected to be seen with the New Horizons data if they were present. Conservatively, a secondary crater and distinctive secondary morphology (such as v-shaped ejecta ) as small as ~0.8 km (~10 pixels) across could be seen in the highest resolution, closest approach strip, which would be produced by an ~16-km-diameter crater nearby (if the largest secondary craters are ~5% of the parent crater size (*Melosh,* 1989; *Singer et al.,* 2013; *Singer et al.,* 2020b), or a larger, more distant crater. There are several craters with $D \gtrsim 16$ km in the neighborhood of the highest resolution strip, but very few fresh craters that have the best chance of exhibiting visible ejecta (not yet hidden by mantling and erosion). Although Pluto and Charon were predicted to have low secondary crater densities given the low primary impact speeds and the bodies' relatively low escape velocities (*Bierhaus and Dones,* 2015; also discussed in Section 4.6), it is still somewhat surprising to find no hint of them.

The 10 largest primary craters on Pluto's encounter hemisphere (excepting the Sputnik basin) range from 60–240 km in diameter. The largest secondary craters from these could be $D \sim 3$–12 km, which would be resolvable over much of Pluto. These largest craters all appear to be relatively old and degraded on Pluto, with no clear signs of ejecta or radial scouring left. There are no clustered craters (or craters with secondary morphology) of the appropriate secondary crater sizes ($D \sim 3$-12 km) around these 10 largest primary craters, or anywhere else, on the encounter hemisphere. On Charon, the largest primary craters are in a similar range to Pluto, with $D \sim 55$–230 km and some of their secondaries should also be visible if they exist on the encounter hemisphere. For Vulcan Planitia, where the lighting is optimal, the 5 largest primary craters are $D \sim 30$–65 km and could produce secondary craters as large as $D \sim 1.5$–3 km (resolvable in the mid-to-high resolution datasets). Although we do see more proximal ejecta deposits on Charon, the secondary craters on Pluto and Charon appear to be erased (for older craters), not visible at our current resolution (as most young/fresh craters are smaller), and/or are not produced in identifiable numbers to begin with. This issue is worthy of future study from a cratering mechanics standpoint.

## 5.2. Terrain Ages

*5.2.1. Summary of Terrain Ages* Pluto displays a wide variety of terrain ages, including several types of young surfaces, a few middle-aged regions, and several heavily-cratered, ancient surfaces (**Fig. 5.2**). Pluto's terrains with high crater densities appear to be quite old, as their crater densities are above the 4 Gyr prediction for the *Greenstreet et al.* (2015) "knee" model (see additional discussion below).

All regions on Charon appear relatively old (~4 Gyr or older; **Fig. 5.3**). Although Vulcan Planitia was resurfaced creating the relatively smooth plain (*Stern et al.,* 2015; *Moore et al.,* 2016; *Beyer et al.,* 2019; *Robbins et al.,* 2019), the plain still has a relatively large number of craters overall, indicating this resurfacing occurred early in Charon's history. Crater densities for Oz Terra and Vulcan Planitia are similar for the small range of crater diameters where they overlap, thus small differences in age cannot be distinguished with the current crater data. We include some information about Charon here for comparison to Pluto, and additional discussion about Charon's terrain ages can be found in the chapter by *Spencer et al.* in this volume.

*5.2.2. Pluto Older Terrains* Many of Pluto's terrains have relatively high crater densities (**Fig. 5.2a**) and are consistent with a surface age estimate of 4 Gyr or older. These older terrains show some spread in their crater densities, consistent with Pluto's complex geologic history. All of the terrains discussed here and the Burney basin are presumed to post-date the Sputnik-basin-forming impact event, which likely created a large resurfaced area on the encounter hemisphere of Pluto from both the topographic feature of the basin itself and its ejecta. There are a few other stratigraphic relationships we can infer. As described above in Section 3.1.3, dark/bright layers are seen in the walls of craters that superpose Burney, implying these alternating layers formed prior to the superposing craters. Because the superposing crater spatial density inside Burney is high, this suggests both Burney and these layers are relatively ancient (again 4 Gyr or older). Extensive descriptions of these regions and their geologic context can be found in Section 3 and also in White et al., this volume.

*5.2.3. Pluto Young and Middle-aged Terrains* Eastern Tombaugh Regio (the eastern side of Pluto's bright heart-shaped region; **Fig. 3.5a,b**) is a middle aged terrain (*Moore et al.*, 2016) with a few large, but eroded, craters, giving a possible upper age limit of a few billion years (**Fig. 5.2b**). Eastern Tombaugh Regio is likely resurfaced by an *ongoing* combination of volatile sublimation and deposition (*Moore et al.*, 2017), which appears more active in the middle portions of the terrain, thus it is difficult to tie this crater retention age to a single geologic event.

Sputnik Planitia has no obvious craters, in either the ~315 m px$^{-1}$ MVIC mosaic covering the entirety of the feature, or any of the higher resolution strips crossing it (down to a pixels scale of ~80 m px$^{-1}$). We can estimate an approximate, upper-limit crater retention age for Sputnik Planitia by calculating the R-value for the hypothetical scenario where one crater exists somewhere on the planitia just below the resolution limit of the images. This implies a surface model age of ~30-50 Myr for the revised power law including the break at $D$ ~13 km to a shallow slope, if 1 crater ~1.6 km across is hidden below the resolution limit (~5 pixels in the 315 m px$^{-1}$ PEMV_P_MVIC_LORRI_CA scan). The polygonal features found there are likely created by sluggish lid convection (*Stern et al.*, 2015; *Moore et al.*, 2016), which could resurface on timescales of roughly 500,000 years, limiting the age of the surface.

Other young areas with few, if any, craters are the chaotic mountain blocks in Sputnik Plantia, the putative cryovolcanic construct Wright Mons to the south west of Sputnik Planitia, and Pluto's "bladed terrain" to the east of Tombaugh Regio. The mountain blocks and Wright Mons share the same $D$ = 1.6 km crater size constraint as Sputnik Planitia but cover a smaller area, thus we derive an older maximum age estimates for these terrains.

The chaotic mountain blocks (*Moore et al.*, 2016; *Skjetne et al.*, 2020) have no obvious craters at a pixel scale of 315 m px$^{-1}$ (the pixel scale available for all mountainous regions combined). Although they do not represent a continuous surface, there are some flat block tops that could host pre- or post-disruption craters, and craters that are large enough could still form and be visible on the uneven terrain. In the highest resolution images (pixel scale of 77 m px$^{-1}$) that cover one region of chaotic blocks, there are some small ($D$ < 1 km) circular or sub-circular features that may be small impacts and one or two larger features that may be craters. With all of the disruption, however, it is difficult to tell between a remnant impact crater and a later collapse feature (e.g., the alcove of a mass wasting event can have a curved upper scarp). Additionally, the morphology of any post-disruption craters could be affected by formation on the many slopes or uneven terrains that make up the chaotic blocks, making them more difficult to identify. In this work we use the absence of clear craters at 315 m px$^{-1}$ for an upper limit, because large post-disruption craters could, in principle, still be visible. This gives an upper limit age of ~200-300 Myr for the mountain blocks (top of the error bar), however, it is also not entirely clear what event is being dated. The block material generally resembles the nearby terrain just outside of Sputnik Planitia (that is not broken up). Thus, this is likely an older surface/crust that was emplaced as part of the Sputnik basin ejecta deposits, with a partial resurfacing from the disruption into chaotic blocks.

An upper limit for crater spatial density of the bladed terrain, from the lack of any distinct craters 2.5 km across or larger (~5 pixels in the 480 m px$^{-1}$ PEMV_P_ MPAN1 scan), is plotted in **Fig. 5.2b.** This yields upper limit ages of ~200-300 Myr old for the bladed terrain. See additional discussion in the chapter by *White et al.*, in this volume.

With no craters on the main topographic mound of Wright Mons, this yields a 1-2 Gyr upper age limit for the surface (if using the top of the error bar). If the one possible 5.5-km-diameter feature is indeed a crater on Wright Mons (see discussion in Section 3.1.4), it would yield an age closer to 3-4 Gyr. These relatively high upper limits on Wright Mons' age do not mean that the feature itself is that old or that the process that built it necessarily only operated early in Pluto's history. First, these are extreme upper limits based on the image resolutions available and the top of the error bar. Second, because the number of craters is divided by the area in order to calculate crater densities, the area of measurement also matters. For two regions, both with no craters, the larger area will give a younger age. This can be seen in **Fig. 5.2b** where the points for the three terrains of Sputnik Planitia, the chaotic mountain blocks, and Wright Mons stack on top of each other. The image resolution is the same, each region just covers a different area. Another way to understand this is that it is statistically less likely for no craters to form over time in a larger area than a smaller one. Wright Mons covers a smaller area than some other regions measured at the same image resolution, thus we cannot constrain the age as well with crater measurements. Work on other geologic indicators in the future may give a better estimate of a minimum age, or at least more realistic "most likely" age, for Wright mons.

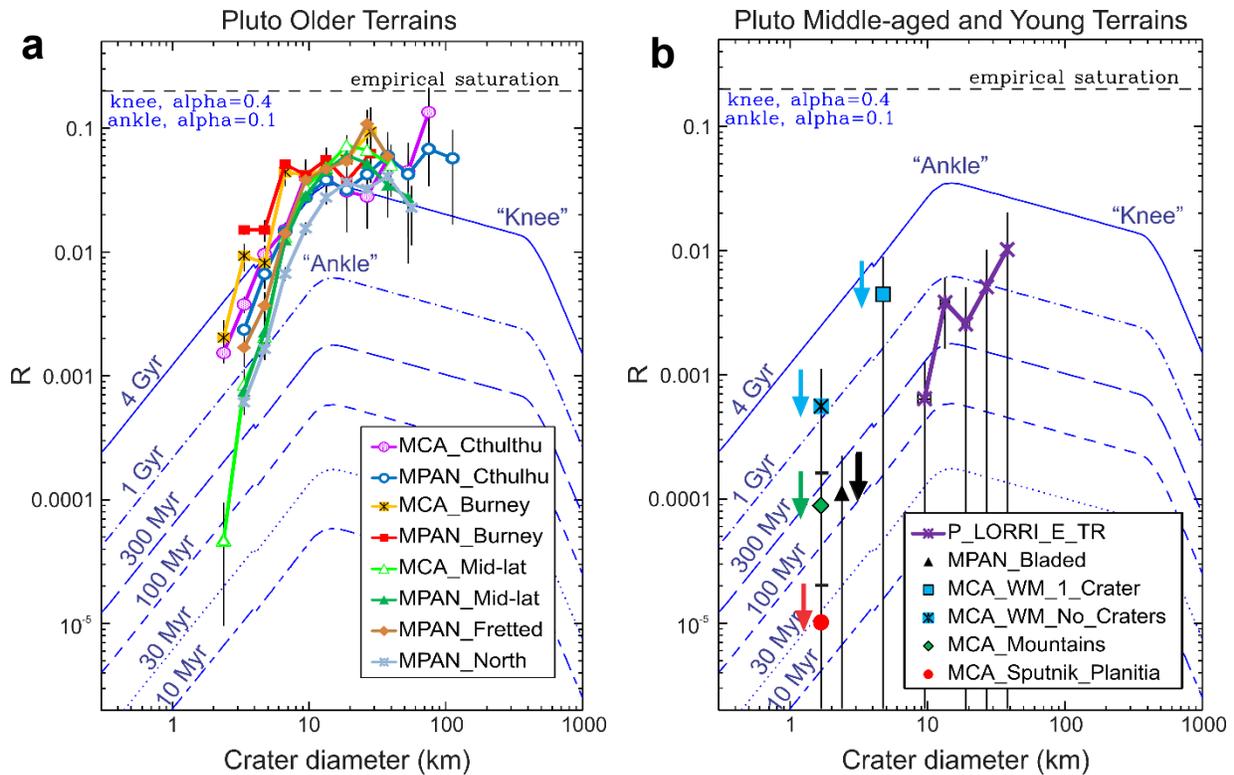

**Figure 5.2** A comparison of crater densities with predictions modified from *Greenstreet et al.* (2015) for (a) Pluto's older terrains, and (b) Pluto's younger and middle-aged terrains. Modified prediction curves include the additional "elbow" break in slope matching Charon's crater distribution (see text for more information). In (b) the single points with arrows next to them represent upper limit ages for Pluto's younger terrains, which display few if any craters. The top of the arrow is placed at the top of the one sigma error bar for each point. For the legends, each dataset is identified by both the image/scan craters were mapped on and the geologic unit (see **Fig. 3.1**): MCA = MVIC Closest Approach scan, MPAN = MVIC PANframe scan, P_LORRI = Pluto LORRI encounter hemisphere mosaic, E_TR = Eastern Tombaugh Regio, and WM = Wright Mons. Additional details about each dataset and the outlines for each region are given in *Singer et al.*, (2019; see main text and supplement). The empirical saturation line is explained in **Fig. 4.2.**

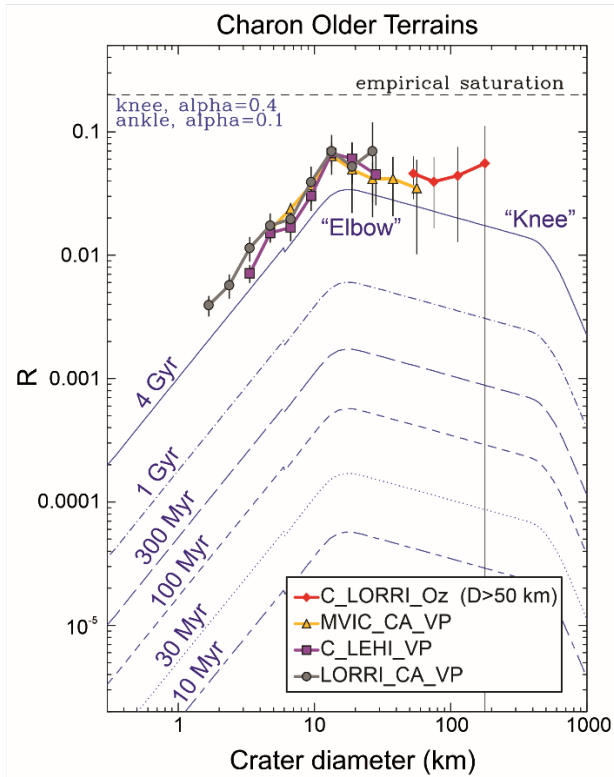

**Figure 5.3** A comparison of crater densities with predictions modified from Greenstreet et al. (2015, 2016) for Charon's terrains. For the legends: C_LORRI = Charon LORRI encounter hemisphere mosaic, MVIC_CA = MVIC Closest Approach scan, C_LEHI = Charon LORRI mosaic associated with the LEISA instrument high resolution scan, LORRI_CA = LORRI closest approach high resolution mosaic, Oz = Oz Terra, and VP = Vulcan Planitia. Additional details about each dataset and the outlines for each region are given in Sections 1 and 3, and in *Singer et al.*, 2019 (see main text and supplement).

## 6. SUMMARY AND FUTURE WORK

The New Horizons flyby of Pluto and Charon provided the first detailed surface images of Kuiper belt objects that are still residing in the Kuiper belt today. This provides unique information about the population of smaller Kuiper belt objects that impacted Pluto and Charon to make craters. We find a range of older, middle aged, and younger terrains on Pluto. The oldest terrains on Pluto have modeled crater retention ages of ~4 Gyr or older. For the youngest terrains on Pluto, those with no identifiable craters, we can only put a rough upper limit on the model ages. The very youngest terrains such as Sputnik Planitia are likely continually resurfaced into the present. On Charon, we find both Oz Terra and Vulcan Planitia have fairly high crater densities that imply relatively old crater retention ages, ~4 Gyr or older. The craters in the Pluto-system also showed that the Kuiper belt has a different size-frequency distribution shape than the asteroid belt, with significantly fewer small ($D$ <10-5 km) objects.

For Pluto, one overarching topic for future research is continued use of crater densities and morphologies along with geologic mapping and stratigraphic relationships to better understand Pluto's geologic history. We have identified many of the major processes operating on Pluto, and they encompass a wide range of process types, from tectonism to volcanism to sublimation/deposition features. Can we quantify the extent of erosion, mantling, or infilling of Pluto's craters? Future classification, measurements, and modeling of geologic processes will all bear on this topic. Additionally, modeling of crater mechanics necessary to reproduce the depth-to-diameter trends and other morphological aspects of craters on Pluto and Charon (including the lack of obvious secondary craters) may also produce insights into surface and sub-surface material properties on Pluto. The information regarding small Kuiper belt objects from New Horizons can also be used with updated cratering rate models to improve estimates of terrain age. For Charon, some areas of lower crater spatial density exist across the plain. These areas are conspicuous to the eye, but further statistical analysis is needed to determine if these low-density areas could be stochastic or if they may indicate some later resurfacing after the majority of Vulcan Planitia was emplaced.

Continued comparison of the Pluto and Charon data to the crater populations of the moons of Jupiter, Saturn, Uranus, and Neptune can provide information on those impactor populations and help constrain which components of the satellite crater populations came from heliocentric vs planetocentric impactors. And finally, comparing what we have learned about small Kuiper belt objects directly to other small body populations, such as the asteroid belt, Centaurs (objects in the giant planet region), or comet populations can inform us on how these bodies form initially, and evolve over the history of the solar system.